\documentclass[%
reprint,
groupedaddress,
nofootinbib,
 amsmath,amssymb,
 aps,
pra,
floatfix,
longbibliography
]{revtex4-1}

\usepackage{graphicx}
\usepackage{dcolumn}
\usepackage{bm}
\usepackage{braket}
\usepackage{physics}
\usepackage{comment}
\usepackage{color}
\usepackage[version=3]{mhchem}
\usepackage{ulem}

\newcommand{\red}[1]{{\textcolor{red}{#1}}}

\usepackage[whole]{bxcjkjatype} 
\usepackage[colorlinks=true,urlcolor=blue,citecolor=blue,linkcolor=blue,breaklinks=true]{hyperref}

\begin{document}

\preprint{APS/123-QED}

\title{Topological phases of electrons induced by electron-magnon interactions}

\author{Kosuke Fujiwara}
\affiliation{Department of Applied Physics, The University of Tokyo, Hongo, Tokyo, 113-8656, Japan}

\author{Takahiro~Morimoto}
\affiliation{Department of Applied Physics, The University of Tokyo, Hongo, Tokyo, 113-8656, Japan}

\date{\today}

\begin{abstract}
Topological phases of electrons such as topological insulators and quantum Hall states typically require strong spin-orbit coupling or magnetic fields.
In this study, we consider an electron system coupled to a spin system, where electrons interact with magnons, quasiparticles of spin waves.
We show that the interaction between electrons and magnons transfers the effect of symmetry breaking in the spin system to the electron system, whereby a non-trivial topological phase can be induced in the electron system that is otherwise topologically trivial.
Through this ``topology transfer'' mechanism, we demonstrate the realization of various topological phases, including quantum Hall and quantum spin Hall insulators, in simple ferromagnetic spin systems, without requiring strong spin-orbit coupling or external magnetic field for electron systems.
\end{abstract}
\maketitle



\section{Introduction}\label{sec:intro}
In recent years, the topological phases of matter in electron systems have attracted much attention \cite{Qi2011TopologicalSuperconductors}. For example, a quantum Hall system with broken time-reversal symmetry (TRS) is characterized by the Chern number \cite{Thouless1982QuantizedPotential,Kohmoto1985TopologicalConductance}, and a chiral edge state appears depending on its value \cite{Hatsugai1993ChernEffect,Hatsugai1993EdgeFunction}. 
In two-dimensional systems with TRS, quantum spin Hall states and topological insulators can appear, which are characterized by the $Z_2$ index \cite{Kane2005QuantumGraphene,Kane2005Z2Effect}. Previous studies have mainly focused on spin-orbit coupling (SOC) as a driving force for realizing these topological phases \cite{Haldane1988ModelAnomaly,Bernevig2006QuantumWells,Konig2007QuantumWells}.
Recently, it has been suggested that in an electron system coupled to a localized spin system, the scalar spin chirality of the spin system can act as an effective magnetic field to realize a quantum Hall system \cite{Hamamoto2015QuantizedCrystal}. Most of the previous studies on the coupling between electron systems and spin systems have treated spins classically and have not considered quantum effects of the spin systems.

Meanwhile, the study of topological phases in magnon systems—the quanta of spin waves—has advanced significantly~\cite{Chumak2015MagnonSpintronics}. 
These studies reveal that the Dzyaloshinskii-Moriya (DM) interaction can break the TRS of the magnon Hamiltonian and give rise to a topological magnon even when the ground state is a collinear ferromagnetic or antiferromagnetic state in which the scalar spin chirality of the classical spins is zero~\cite{Owerre2016AInsulator,Kim2016RealizationSpins,Bhowmick2020AntichiralLattice,Bhowmick2023TuningMagnons,Kim2022TopologicalBreaking,Kawano2019ThermalAntiferromagnet}.
In particular, the thermal Hall effect, which is an analog of the quantum Hall effect in magnon systems, has been proposed theoretically~\cite{Katsura2010TheoryMagnets,Matsumoto2011TheoreticalFerromagnets,Matsumoto2011RotationalEffect} and confirmed experimentally~\cite{Onose2010ObservationEffect}. The magnon spin Nernst effect~\cite{Zyuzin2016MagnonAntiferromagnets,Cheng2016SpinAntiferromagnets}, an analog of the quantum spin Hall effect, and the $Z_2$ index in magnon systems~\cite{Kondo2019Z2Systems,Kondo2019Three-dimensionalSystems} have also been studied. The effects of magnons on electron systems have been investigated in terms of magnon-mediated superconductivity ~\cite{Kargarian2016AmpereanInsulators,Rohling2018SuperconductivityMagnons,Mland2023TopologicalMagnons,VinasBostrom2024Magnon-mediatedWire} and the lifetime of electrons~\cite{Mland2021Electron-magnonInsulator,Jo2021VisualizingFerromagnet,Sourounis2025Interaction-inducedSemimetals}. However, the effects of DM interactions and topological magnons on electron systems have not been fully investigated.

In this paper, we focus on the interaction between the spin system and the electron system, and study how TRS breaking in the magnon Hamiltonian affects the electron system theoretically. We demonstrate that the TRS of the electron system is broken due to the electron-magnon interaction and that the topology of the electron system can change even when it is initially in a topologically trivial state.

\begin{figure}[tb]
    \centering
    \includegraphics[width=0.9\linewidth]{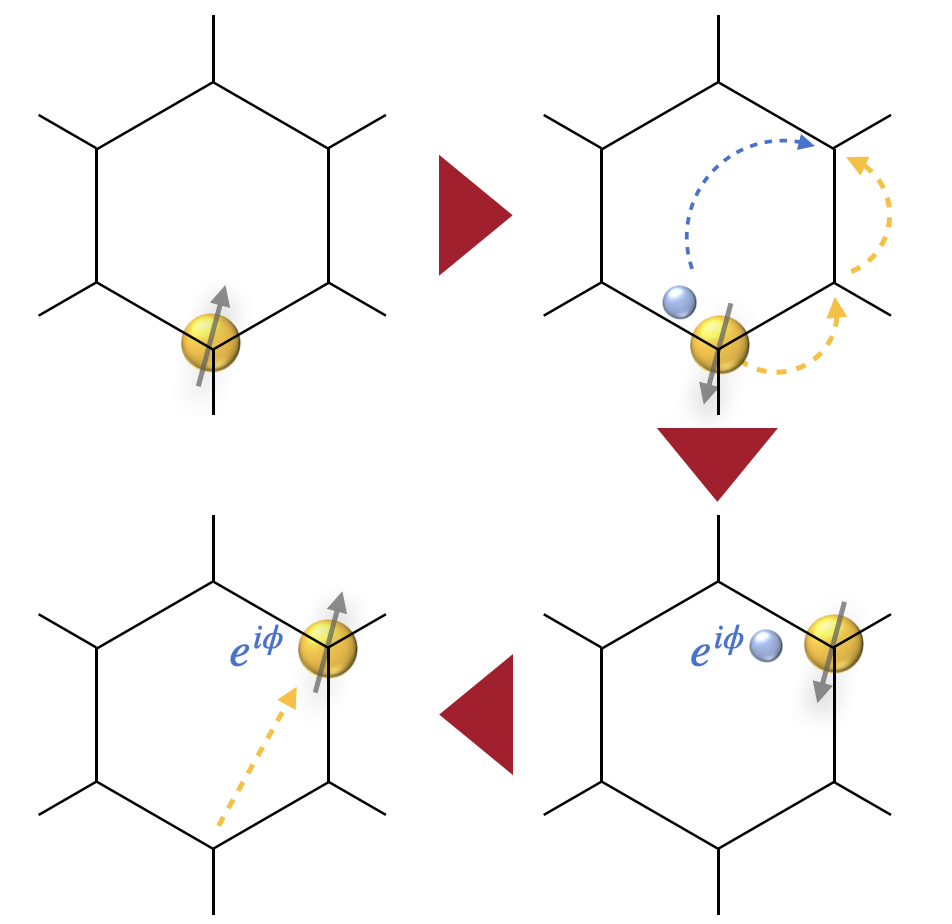}
    \caption{Schematic of magnon-assisted electron hopping leading to a topological electronic phase. Hoppings of electrons (yellow balls) and magnons (blue balls) are shown in real space. Focusing on the initial state and final state, the electron hops to the next-nearest-neighbor site with acquiring the phase factor.} 
    \label{fig:schematic}
\end{figure}

The underlying mechanism can be understood through a heuristic real-space picture, as illustrated in Fig.~\ref{fig:schematic} for an electron and spin system on the honeycomb lattice. The process is effectively a second-order virtual transition for an electron hopping between next-nearest-neighbor sites. First, at an initial site, the electron-magnon interaction causes an electron to flip its spin while creating or annihilating a magnon, entering a virtual intermediate state. While an electron's hopping typically acquires no phase without SOC, the magnon's hopping between next-nearest-neighbor sites can acquire a non-trivial phase due to the DM interaction. Finally, at the destination site, a second electron-magnon interaction brings the system back to its original low-energy manifold. The net effect of this process is an effective next-nearest-neighbor hopping for the electron that has acquired a complex phase factor. This mechanism is analogous to the complex hopping in the Haldane model, where the phase is generated not by an external magnetic field but by the internal properties of the coupled magnon system.

In the following sections, we evaluate electron-magnon interactions and show that topological magnons can induce topological phases in electron systems.
In section~\ref{sec:methods}, we present general methods to incorporate electron-magnon interactions as a self-energy of electrons. In section~\ref{sec:QH}, we apply these methods to a heterostructure tight-binding model on the honeycomb lattice. We show that the interaction between topological magnons and electrons can open a gap in the electron band and that the electron system can become a quantum Hall insulator. In section~\ref{sec:QSHI}, from the analogy with the Kane-Mele model~\cite{Kane2005QuantumGraphene,Kane2005Z2Effect}, which is created by stacking two Haldane models, we show that a quantum spin Hall system can be created by stacking time-reversed pairs of spin systems. Furthermore, we show that in systems where spin is not conserved, an effective SOC arises from electron-magnon interactions. In section~\ref{sec:discussion}, we discuss the scope of our theory and candidate materials.

\section{Methods}\label{sec:methods}
We consider a Hamiltonian consisting of an electron system and a spin system
\begin{equation}
    \mathcal{H}_{\text{tot}} = \mathcal{H}_{\text{e}} + \mathcal{H}_{\text{s}} + \mathcal{H}_{\text{int}},
\end{equation}
where $\mathcal{H}_{\text{e}}$ is the Hamiltonian of non-interacting electrons
\begin{equation}
    \mathcal{H}_{\text{e}}=\sum_{\vb*{k}}\sum_{\alpha,\beta}\phi^\dagger_{e,\vb*{k},\alpha}H_{\text{e},\vb*{k}}^{\alpha,\beta}\phi_{e,\vb*{k},\beta},
\end{equation}
\begin{equation}
    \phi_{e,\vb*{k}}=(c_{1,\uparrow,\vb*{k}},\cdots,c_{n,\uparrow,\vb*{k}},c_{1,\downarrow,\vb*{k}},\cdots,c_{n,\downarrow,\vb*{k}})^T,
\end{equation}
where $c_{i,\mu,\vb*{k}}$ is the annihilation operator of the electron at $i$-th site in a unit cell with spin $\mu$ and $n$ is the number of sites in the unit cell. $\mathcal{H}_{\text{s}}$ is the Hamiltonian of the spin system which has an ordered ground state and $\mathcal{H}_{\text{int}}$ is the interaction between itinerant electrons and local spins given by
\begin{equation}
    \mathcal{H}_{\text{int}} = \sum_{i,\mu,\nu}-J_{ex}\vb*{S}_{i}\cdot c_{i\mu}^\dagger\vb*{\sigma}_{\mu\nu}c_{i\nu},\label{eq:interaction}
\end{equation}
where $S_i$ is the spin operator at site $i$, $\vb*{\sigma}$ is the Pauli matrix, and $J_{ex}$ is the exchange coupling constant. This interaction is realized by, for instance, the Hund's coupling in a magnetic metal or the proximity-induced exchange interaction in a magnet/metal heterostructure.

By rotating the spin quantization axis along the spin structure, we can utilize the Holstein-Primakoff transformation~\cite{Holstein1940FieldFerromagnet} to expand the spin Hamiltonian $\mathcal{H}_{\text{s}}$ and the interaction $\mathcal{H}_{\text{int}}$ in terms of magnon operators. In particular, we consider $\vb*{S}_i=\mathcal{R}_i\tilde{\vb*{S}}_i$ where $\mathcal{R}_i$ is the rotation matrix and $\tilde{\vb*{S}}_i=(0,0,S)^T$ is the spin vector in the rotated spin coordinate. Then, the Holstein-Primakoff transformation is given by
\begin{subequations}
    \begin{align}
        \tilde{S}_{j}^+ &\sim \sqrt{2S}a_j,\\
        \tilde{S}_{j}^- &\sim \sqrt{2S}a_j^\dagger,\\
        \tilde{S}_{j}^z &= S-a_j^\dagger a_j,
    \end{align}
\end{subequations}
where $a_j$ is the annihilation operator of the magnon at $j$-th site. Applying the Holstein-Primakoff transformation and the Fourier transformation to the spin Hamiltonian $\mathcal{H}_{\text{s}}$, the magnon Hamiltonian is obtained in the form of the following bosonic Bogoliubov-de Gennes (BdG) Hamiltonian
\begin{subequations}
\begin{align}
    &\mathcal{H}_{\text{m}} = \sum_{\vb*{k}}\phi^\dagger_{m,\vb*{k}}H_{\text{m},\vb*{k}}\phi_{m,\vb*{k}}\\
    &\phi_{m,\vb*{k}}=(a_{1,\vb*{k}},\cdots,a_{n,\vb*{k}},a^\dagger_{1,-\vb*{k}},\cdots,a^\dagger_{n,-\vb*{k}})^T.
\end{align}
\end{subequations}
The interaction Hamiltonian is also expanded by the magnon as
\begin{equation}
    \mathcal{H}_{\text{int}} = \mathcal{H}_{\text{int}0}+\mathcal{H}_{\text{int}1}+\mathcal{H}_{\text{int}2},
\end{equation}
where
\begin{align}
    \mathcal{H}_{\text{int}0}=&\sum_{\vb*{k}}\sum_{l}\sum_{\mu}H_{\text{int}0}^{l,\mu}c^\dagger_{l,\mu,\vb*{k}}c_{l,\mu,\vb*{k}},\\
    \mathcal{H}_{\text{int}1}=&\frac{1}{\sqrt{N}}\sum_{\vb*{k},\vb*{q}}\sum_{l}\sum_{\mu,\nu}[H^{l,\mu\nu}_{\text{int}1}c^\dagger_{l,\mu,\vb*{k}}c_{l,\nu,\vb*{q}}a_{\vb*{k}-\vb*{q},l}+h.c.],\\
    \mathcal{H}_{\text{int}2}=&\frac{1}{N}\sum_{\vb*{k},\vb*{q},\vb*{p}}\sum_{l}\sum_{\mu,\nu}H^{l,\mu\nu}_{\text{int}2}c^\dagger_{l,\mu,\vb*{k}}c_{l,\nu,\vb*{q}}a^\dagger_{\vb*{p},l}a_{\vb*{k}-\vb*{q}+\vb*{p},l}.
\end{align}
Here, $l$ represents the degrees of freedom of the sublattice, ranging from $1$ to $n$, and $N$ is the number of unit cells.
The interaction Hamiltonian $\mathcal{H}_{\text{int}0}$ describes the interaction between electrons and classical spins and corresponds to the effective magnetic field. $\mathcal{H}_{\text{int}1}$ and $\mathcal{H}_{\text{int}2}$ incorporate the quantum effects and describe interaction between electrons and magnons. 
In particular, $\mathcal{H}_{\text{int}2}$ corresponds to the correction of spin expectation value $S\rightarrow \Braket{\tilde{S}^z}=S-\Braket{a^\dagger a}$.
By using $\phi_e$ and $\phi_m$, $\mathcal{H}_{\text{int}1}$ and $\mathcal{H}_{\text{int}2}$ can be written as
\begin{subequations}
\begin{align}
    \mathcal{H}_{\text{int}1}=&\frac{1}{\sqrt{N}}\sum_{\vb*{k},\vb*{q}}\sum_{\alpha,\beta,\gamma}V^{\alpha\beta\gamma}\phi^\dagger_{e,\vb*{k},\alpha}\phi_{e,\vb*{q},\beta}\phi_{m,\vb*{k}-\vb*{q},\gamma}\\
    \mathcal{H}_{\text{int}2}=&\frac{1}{N}\sum_{\vb*{k},\vb*{q},\vb*{p}}\sum_{\alpha,\beta,\gamma}W^{\alpha\beta\gamma}\phi^\dagger_{e,\vb*{k},\alpha}\phi_{e,\vb*{q},\beta}\phi^\dagger_{m,\vb*{p},\gamma}\phi_{m,\vb*{k}-\vb*{q}+\vb*{p},\gamma}.
\end{align}
\end{subequations}
Due to the hermicity of interactions, $V$ and $W$ satisfy
\begin{subequations}
\begin{align}
    V^{\alpha\beta\gamma}&=\sum_{\gamma^\prime}\sigma_{1,\gamma\gamma^\prime}(V^{\beta\alpha\gamma^\prime})^*\\
    W^{\alpha\beta\gamma}&=(W^{\beta\alpha\gamma})^*,
\end{align}
where $\sigma_1=\sigma_x\otimes 1_n$.
\end{subequations}
Therefore, the total Hamiltonian is
\begin{equation}
    \mathcal{H}_{\text{tot}} = \mathcal{H}^\prime_{\text{e}} + \mathcal{H}_{\text{m}} + \mathcal{H}_{\text{int}1} + \mathcal{H}_{\text{int}2},
\end{equation}
where $\mathcal{H}^\prime_{\text{e}}$ is the electron Hamiltonian with the interaction between the electron and the classical spin as
\begin{equation}
    \mathcal{H}^\prime_{\text{e}} = \mathcal{H}_{\text{e}} + \mathcal{H}_{\text{int}0}.
\end{equation}

Let us consider bases $\psi_{e,\vb*{k}}$ and $\psi_{m,\vb*{k}}$ which diagonalize the Hamiltonian $H^\prime_{\text{e},\vb*{k}}$ and $H_{\text{m},\vb*{k}}$. By using these bases, we can write the total Hamiltonian as
\begin{align}
    \mathcal{H}_{\text{tot}}=&\sum_{\vb*{k}}\sum_{\alpha}[\psi^\dagger_{e,\vb*{k},\alpha}\varepsilon_{\vb*{k},\alpha}\psi_{e,\vb*{k},\alpha}+\psi^\dagger_{m,\vb*{k},\alpha}\omega_{\vb*{k},\alpha}\psi_{m,\vb*{k},\alpha}]\notag\\
    &+\frac{1}{\sqrt{N}}\sum_{\vb*{k},\vb*{q}}\sum_{\alpha,\beta,\gamma}\tilde{V}^{\alpha\beta\gamma}_{\vb*{k},\vb*{q}}\psi^\dagger_{e,\vb*{k},\alpha}\psi_{e,\vb*{q},\beta}\psi_{m,\vb*{k}-\vb*{q},\gamma}\notag\\
    &+\frac{1}{N}\sum_{\vb*{k},\vb*{q}}\sum_{\alpha,\beta,\gamma,\delta}\tilde{W}^{\alpha\beta\gamma\delta}_{\vb*{k},\vb*{q},\vb*{p}}\psi^\dagger_{e,\vb*{k},\alpha}\psi_{e,\vb*{q},\beta}\psi^\dagger_{m,\vb*{p},\gamma}\psi_{m,\vb*{k}-\vb*{q}+\vb*{p},\delta}.
\end{align}
The hermiticity of the interaction requires that $\tilde{V}$ and $\tilde{W}$ satisfy
\begin{subequations}
\begin{align}
    &\tilde{V}^{\alpha\beta\gamma}_{\vb*{k},\vb*{q}}=\sum_{\gamma^\prime}\sigma_{1,\gamma\gamma^\prime}(\tilde{V}^{\beta\alpha\gamma^\prime}_{\vb*{q},\vb*{k}})^*,\\
    &\tilde{W}^{\alpha\beta\gamma\delta}_{\vb*{k},\vb*{q},\vb*{p}}=(\tilde{W}^{\beta\alpha\delta\gamma}_{\vb*{q},\vb*{k},\vb*{k}-\vb*{q}+\vb*{p}})^*=\sum_{\gamma^\prime,\delta^\prime}\sigma_{1,\gamma\gamma^\prime}\sigma_{1,\delta\delta^\prime}(\tilde{W}^{\beta\alpha\gamma^\prime\delta^\prime}_{\vb*{q},\vb*{k},-\vb*{p}})^*.
\end{align}
\end{subequations}

\begin{figure}[tb]
    \centering
    \includegraphics[width=\linewidth]{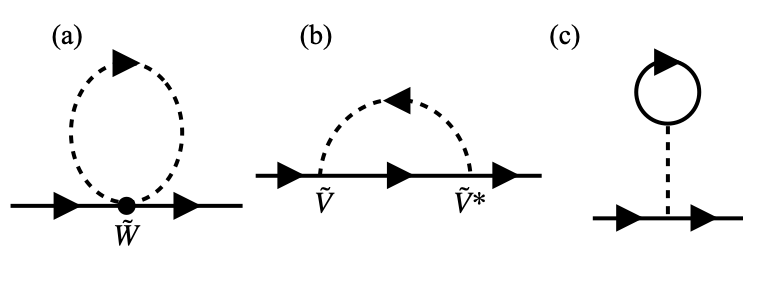}
    \caption{Feynman diagrams for the lowest-order electron self-energy contributions: (a) The correction  term $\Sigma_{cor}$ [Eq.~\eqref{General_Sigma_cor}], (b) the scattering term $\Sigma_{sca}$ [Eq.~\eqref{General_Sigma_sca}], and (c) the tadpole term $\Sigma_{tad}$ [Eq.~\eqref{General_Sigma_tad}].} 
    \label{fig:Diagram}
\end{figure}

Now, we incorporate the effect of magnon to the electron system as the self-energy. We consider the lowest order perturbation with respect to the interactions $\mathcal{H}_{int1}$ and $\mathcal{H}_{int2}$. The Green's function of electron $\mathcal{G}_e$ in the imaginary time formalism is given by 
\begin{align}
    &\mathcal{G}_{e,k,\alpha\beta}(\tau)=\notag\\
    &\mathcal{G}_{e,k,\alpha\beta}^0(\tau)+\int^\beta_0d\tau_1\langle T_\tau \mathcal{H}_{\text{int}2}(\tau_1)\psi_{e,\vb*{k},\alpha}(\tau)\psi^\dagger_{e,\vb*{k},\beta}\rangle\notag\\
    &-\frac{1}{2}\int^\beta_0d\tau_1\int^\beta_0d\tau_2\langle T_\tau \mathcal{H}_{\text{int}1}(\tau_1)\mathcal{H}_{\text{int}1}(\tau_2)\psi_{e,\vb*{k},\alpha}(\tau)\psi^\dagger_{e,\vb*{k},\beta}\rangle,\label{eq:green's_func}
\end{align}
where $\mathcal{G}^0$ is the unperturbed Green's function $\mathcal{G}^0_{e,\vb*{k},\alpha\beta}(\tau)=-\langle T_\tau \psi_{e,\vb*{k},\alpha}(\tau)\psi^\dagger_{e,\vb*{k},\beta}\rangle$, $\beta=1/k_BT$ is the inverse temperature, $T_\tau$ represents the imaginary time ordering of operators, and $\langle\cdots\rangle$ is the thermal average with respect to the unperturbed Hamiltonian $\mathcal{H}_e^\prime$. The corresponding Feynman diagrams are shown in Fig.~\ref{fig:Diagram}.

The contribution of $\mathcal{H}_{\text{int}2}$ to the self-energy is written as
\begin{align}
    \Sigma^{\alpha\beta}_{cor,\vb*{k},\omega}= \frac{1}{N}\sum_{\vb*{q}}\sum_{\gamma} \tilde{W}_{\vb*{k},\vb*{k},\vb*{q}}^{\alpha\beta\gamma\gamma}\sigma_{3,\gamma\gamma}n_B(\sigma_{3,\gamma\gamma}\omega_{\vb*{q},\gamma}),\label{General_Sigma_cor}
\end{align}
where $n_B$ is the Bose-Einstein distribution function and $\sigma_3$ is $\sigma_z\otimes 1_n$. This term originates from $H_{\text{int}2}$ and thus corresponds to the correction of the spin expectation value $S\rightarrow \Braket{\tilde{S}^z}$, due to fluctuations.
The contribution of $\mathcal{H}_{\text{int}1}$ to the self-energy is divided into two terms, $\Sigma_{sca}$ and $\Sigma_{tad}$. $\Sigma_{sca}$ is written as
\begin{align}
    \Sigma^{\alpha\beta}_{sca,\vb*{k},\omega}&= \frac{1}{2N}\sum_{\vb*{q}}\sum_{\gamma,\delta}\tilde{V}_{\vb*{q},\vb*{k}}^{\gamma\alpha\delta}(\tilde{V}_{\vb*{q},\vb*{k}}^{\gamma\beta\delta})^*\notag\\
    &\times\frac{\sigma_{3,\delta\delta}(f(\varepsilon_{\vb*{q},\gamma})+n_B(\sigma_{3,\delta\delta}\omega_{\vb*{q}-\vb*{k},\delta}))}{\omega-\varepsilon_{\vb*{q},\gamma}+\sigma_{3,\delta\delta}\omega_{\vb*{q}-\vb*{k},\delta}+i\eta},\label{General_Sigma_sca}
\end{align}
where $f$ is the Fermi-Dirac distribution function (See Appendix \ref{sec:appendix_derivation}). This term corresponds to the scattering of electrons and magnons.
$\Sigma_{tad}$ is given by
\begin{equation}
    \Sigma_{tad,\vb*{k},\omega}^{\alpha\beta}=-\frac{1}{2N}\sum_{\vb*{q}}\sum_{\gamma,\delta}\tilde{V}_{\vb*{k},\vb*{k}}^{\beta\alpha\delta}(\tilde{V}_{\vb*{q},\vb*{q}}^{\gamma\gamma\delta})^*\frac{f(\varepsilon_{\vb*{q},\gamma})}{\omega_{\vb*{0},\delta}}.\label{General_Sigma_tad}
\end{equation}
In our formalism, the self-energy is evaluated by using the unperturbed Green's function. We evaluate the renormalized Green's function of magnons in Appendix \ref{appendix:magnon} and check the validity of using the unperturbed Green's function.

When $\mathcal{H}_{\text{e}}^\prime$ can be decoupled to up spin sector and down spin sector, we can simplify the above equations. 
In this case, we use eigenstates of up or down spin $\psi_{\mu,\vb*{k}}$ and their energy $\varepsilon_{\mu,\vb*{k}}$, instead of the general band eigenstates $\psi_{e,\vb*{k}}$ and their energies $\varepsilon_{\vb*{k}}$. The total Hamiltonian (Eq. (15)) is rewritten as
\begin{align}
    \mathcal{H}_{\text{tot}}=&\sum_{\vb*{k}}\bigg[\sum_{\alpha=1}^n\sum_{\mu}\psi^\dagger_{\mu,\vb*{k},\alpha}\varepsilon_{\mu,\vb*{k},\alpha}\psi_{\mu,\vb*{k},\alpha}\notag\\
    &+\sum_{\gamma=1}^{2n}\psi^\dagger_{m,\vb*{k},\gamma}\omega_{\vb*{k},\gamma}\psi_{m,\vb*{k},\gamma}\bigg]\notag\\
    &+\frac{1}{\sqrt{N}}\sum_{\vb*{k},\vb*{q}}\sum_{\mu,\nu}\sum_{\alpha,\beta=1}^n\sum_{\gamma=1}^{2n}\notag\\
    &\tilde{V}^{\alpha\beta\gamma,\mu\nu}_{\vb*{k},\vb*{q}}\psi^\dagger_{\mu,\vb*{k},\alpha}\psi_{\nu,\vb*{q},\beta}\psi_{m,\vb*{k}-\vb*{q},\gamma}\notag\\
    &+\frac{1}{N}\sum_{\vb*{k},\vb*{q}}\sum_{\mu,\nu}\sum_{\alpha,\beta=1}^n\sum_{\gamma,\delta=1}^{2n}\notag\\
    &\tilde{W}^{\alpha\beta\gamma\delta,\mu\nu}_{\vb*{k},\vb*{q},\vb*{p}}\psi^\dagger_{\mu,\vb*{k},\alpha}\psi_{\nu,\vb*{q},\beta}\psi^\dagger_{m,\vb*{p},\gamma}\psi_{m,\vb*{k}-\vb*{q}+\vb*{p},\delta},
\end{align}
where $\alpha$ and $\beta$ are indices of $\psi_{\mu,\vb*{k}}$ and $\gamma$ and $\delta$ are indices of $\psi_{m,\vb*{k}}$.
The hermiticity of the interactions requires
\begin{align}
    &\tilde{V}^{\alpha\beta\gamma,\mu\nu}_{\vb*{k},\vb*{q}}=\sum_{\gamma^\prime}\sigma_{1,\gamma\gamma^\prime}(\tilde{V}^{\beta\alpha\gamma^\prime,\nu\mu}_{\vb*{q},\vb*{k}})^*,\notag\\
    &\tilde{W}^{\alpha\beta\gamma\delta,\mu\nu}_{\vb*{k},\vb*{q},\vb*{p}}=(\tilde{W}^{\beta\alpha\delta\gamma,\nu\mu}_{\vb*{q},\vb*{k},\vb*{k}-\vb*{q}+\vb*{p}})^*=\sum_{\gamma^\prime,\delta^\prime}\sigma_{1,\gamma\gamma^\prime}\sigma_{1,\delta\delta^\prime}(\tilde{W}^{\beta\alpha\gamma^\prime \delta^\prime,\nu\mu}_{\vb*{q},\vb*{k},-\vb*{p}})^*,
\end{align}
Then the self-energy for the electrons is given by
\begin{subequations}
    \begin{align}
    \Sigma^{\alpha\beta,\mu\nu}_{cor,\vb*{k},\omega}=& \frac{1}{N}\sum_{\vb*{q}}\sum_{\gamma} \tilde{W}_{\vb*{k},\vb*{k},\vb*{q}}^{\alpha\beta\gamma\gamma,\mu\nu}\sigma_{3,\gamma\gamma}n_B(\sigma_{3,\gamma\gamma}\omega_{\vb*{q},\gamma}),\\
    \Sigma^{\alpha\beta,\mu\nu}_{sca,\vb*{k},\omega}=&\frac{1}{2N}\sum_{\vb*{q}}\sum_{\rho}\sum_{\gamma=1}^n\sum_{\delta=1}^{2n} \sigma_{3,\delta\delta}\tilde{V}_{\vb*{q},\vb*{k}}^{\gamma\alpha\delta,\rho\mu}(\tilde{V}_{\vb*{q},\vb*{k}}^{\gamma\beta\delta,\rho\nu})^*\notag\\
    &\times\frac{(f(\varepsilon_{\rho,\vb*{q},\gamma})+n_B(\sigma_{3,\delta\delta}\omega_{\vb*{q}-\vb*{k},\delta}))}{\omega-\varepsilon_{\rho,\vb*{q},\gamma}+\sigma_{3,\delta\delta}\omega_{\vb*{q}-\vb*{k},\delta}+i\eta}\\
    \Sigma_{tad,\vb*{k},\omega}^{\alpha\beta,\mu\nu}=&-\frac{1}{2N}\sum_{\vb*{q}}\sum_{\gamma,\delta}\sum_{\rho}\tilde{V}_{\vb*{k},\vb*{k}}^{\beta\alpha\delta,\nu\mu}(\tilde{V}_{\vb*{q},\vb*{q}}^{\gamma\gamma\delta,\rho\rho})^*\notag\\
    &\times\frac{f(\varepsilon_{\rho,\vb*{q},\gamma})}{\omega_{\vb*{0},\delta}}.
\end{align}\label{eq:self_energy_decoupled}
\end{subequations}

\section{Hall insulator induced by magnon} \label{sec:QH}
In this section, we apply the method described in the previous section to a tight-binding model on the honeycomb lattice and show that we can realize the quantum Hall insulator by the interaction between electrons and topological magnons.

\subsection{Model} \label{subsec:model}
\begin{figure}[htb]
    \centering
    \includegraphics[width=\linewidth]{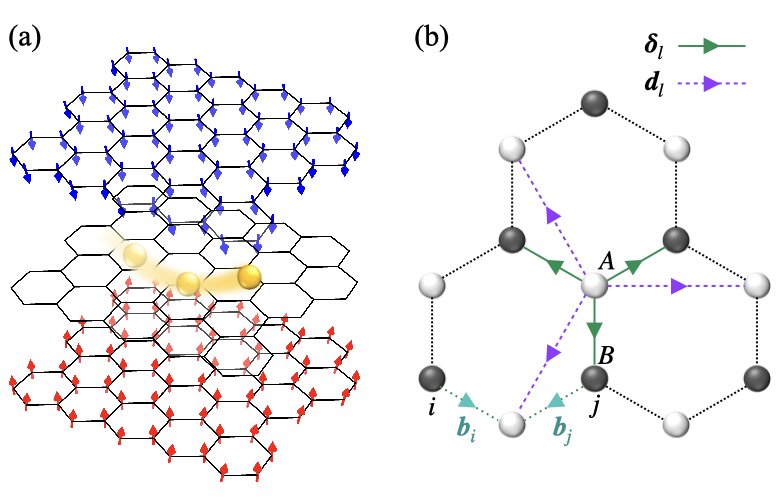}
    \caption{Schematic picture of three-layer honeycomb lattice model.  (a) Electron system sandwiched by top and bottom spin systems. Top and bottom layers are ferromagnetically ordered in an anti-parallel way.
    (b) Honeycomb lattice structure of each layer. Relevant vectors for electron hopping and spin interactions are indicated.} 
    \label{fig:honeycomb}
\end{figure}

We consider a multilayer system as shown in Fig.\ref{fig:honeycomb}. In this system, the bottom and top layers are ferromagnetic spin systems with the DM interaction and the middle layer is an electron system on the honeycomb lattice. In the ground state, spins in the bottom layer are aligned along the $z$ direction and spins in the top layer are aligned along the $-z$ direction as shown in Fig.\ref{fig:honeycomb} (a). Therefore, the exchange field (the interaction of electrons and classical spins) is zero. 

In this system, the Hamiltonian is given by
\begin{equation}
    \mathcal{H}_{\text{tot}} = \mathcal{H}_{\text{e}} + \mathcal{H}_{\text{s},b} + \mathcal{H}_{\text{s},t} + \mathcal{H}_{\text{int}},~\label{eq:total_hamiltonian}
\end{equation}
where $\mathcal{H}_{\text{e}}$ is the electron Hamiltonian, $\mathcal{H}_{\text{s},t}$ is the Hamiltonian of the top layer spin system, $\mathcal{H}_{\text{s},b}$ is the Hamiltonian of the bottom layer spin system, and $\mathcal{H}_{\text{int}}$ is the interaction between electrons and spins in the top and bottom layers.
The tight-binding Hamiltonian of an electron is given by
\begin{align}
    \mathcal{H}_{\text{e}} = \sum_{\mu}\big[&-t_1\sum_{\langle i,j\rangle }(c_{i,\mu}^\dagger c_{j,\mu} + \text{h.c.})+\notag\\
    &-t_2\sum_{\langle\langle i,j\rangle\rangle}(c_{i,\mu}^\dagger c_{j,\mu} + \text{h.c.}) - \varepsilon_F\sum_{i}c_{i,\mu}^\dagger c_{i,\mu}\big],\label{eq:electron}
\end{align}
where $c_{i\mu}^\dagger$ is the electron creation operator at site $i$ with spin $\mu$, $t_1$ and $t_2$ are hopping parameters for nearest neighbor sites and next-nearest-neighbors sites, and $\varepsilon_F$ is the chemical potential. We denote the two sublattices of the honeycomb lattice by $A$, $B$ and using the Fourier transformation, we obtain
\begin{subequations}
    \begin{align}
        &\mathcal{H}_{\text{e}} = \sum_{\vb*{k}}\phi_{e,\vb*{k}}^\dagger H_{\text{e},\vb*{k}}\phi_{e,\vb*{k}},\\
        &\phi_{e,\vb*{k}}=(c_{A,\uparrow,\vb*{k}},c_{B,\uparrow,\vb*{k}},c_{A,\downarrow,\vb*{k}},c_{B,\downarrow,\vb*{k}})^T.
    \end{align}
\end{subequations}
Here, $H_{\text{e},\vb*{k}}$ is
\begin{equation}
    H_{\text{e},\vb*{k}} = \begin{pmatrix}
         H_{\text{e},\vb*{k},\uparrow} & 0\\
        0 &  H_{\text{e},\vb*{k},\downarrow},
        \end{pmatrix},
\end{equation}
with
\begin{equation}
    H_{\text{e},\vb*{k},\uparrow}=H_{\text{e},\vb*{k},\downarrow} = \begin{pmatrix}
        -2t_2g(\vb*{k})-\varepsilon_F & -t_1\gamma(\vb*{k})\\
        -t_1\gamma(\vb*{k})^* & -2t_2g(\vb*{k})-\varepsilon_F,
        \end{pmatrix}, \label{eq:Hek}
\end{equation}
and $\gamma(\vb*{k})=\sum_{l}e^{i\vb*{k}\cdot\delta_l}$ and $g(\vb*{k})=\sum_l\cos{(\vb*{k}\cdot\vb*{d}_l)}$. The nearest neighbor vectors are $\delta_l=(0,-a),(\pm\sqrt{3}a/2,a/2)$ and the next nearest-neighbor vectors are $\vb*{d}_l=(\sqrt{3}a,0),(-\sqrt{3}a/2,\pm3a/2)$.

We consider the ferromagnetic spin system on the honeycomb lattice. The spin Hamiltonian is given by
\begin{align}
    \mathcal{H}_{\text{s}_\lambda} = &\sum_{\langle i,j\rangle}-J_1\vb*{S}_i\cdot \vb*{S}_j-\sum_i \Delta_z(S_i^{z})^2\notag\\
    &+ \sum_{\langle\langle i,j\rangle\rangle}[-J_2S_i\cdot S_j-D\vb*{\xi}_{\lambda,ij}\cdot(\vb*{S}_i\times \vb*{S}_j)]\label{eq:spin_model}
\end{align}
where $S_i$ is the spin operator at site $i$, $J_1$ and $J_2$ are the nearest-neighbor and next-nearest-neighbor exchange interactions, respectively, $\Delta_z$ is the spin axis anisotropy, $D$ is the DM interaction for the next-nearest-neighbor and $\lambda$ denotes the layer index ($\lambda=b$ for the bottom layer and $\lambda=t$ for the top layer). The factor $\xi_{\lambda,ij}$ indicates the sign of the DM interaction; for the bottom layer, it is defined as $\xi_{b,ij}=2/(\sqrt{3}a^2)\vb*{b}_i\times\vb*{b}_j=\pm(0,0,1)$ where $\vb*{b}_i$ and $\vb*{b}_j$ are vectors from site $i$ and site $j$ to the common nearest neighbor lattice respectively as shown in Fig.~\ref{fig:honeycomb} (b).
The spin system for the top layer is obtained by rotating the bottom-layer system by $\pi$ around the $y$ axis. Therefore, the DM interaction has an opposite sign as $\xi_{t,ij}=-\xi_{b,ij}$ in the top layer.
Here, we assume that the ground state of the spin system is ferromagnetic and $\vb*{S}_i=(0,0,S)$ ($\vb*{S}_i=(0,0,-S)$) in the bottom (top) layer. Due to the ferromagnetic order, the spin Hamiltonian breaks the TRS. However, when $D=0$, the Hamiltonian has an effective TRS, defined as the combination of the time-reversal operation and a spin rotation (e.g. a $\pi$ rotation about the $y$ axis in this model). Therefore, DM interaction is the key term that breaks the effective TRS.

We use a Holstein-Primakoff transformation to obtain the magnon Hamiltonian. By expanding the spin Hamiltonian with magnons up to the order of $1/S$, the magnon Hamiltonian is given by
\begin{subequations}
\begin{align}
    &\mathcal{H}_{\text{m}_\lambda}=\frac{1}{2}\sum_{\vb*{k}}\phi^\dagger_{m_\lambda}(\vb*{k})H_{\text{m}_\lambda}(\vb*{k})\phi_{m_\lambda}(\vb*{k})\\
    &\phi_{m_\lambda}(\vb*{k})=(a_{\lambda,A,\vb*{k}},a_{\lambda,B,\vb*{k}},a_{\lambda,A,-\vb*{k}}^\dagger,a_{\lambda,B,-\vb*{k}}^\dagger)^T.
\end{align}
\end{subequations}
Here, operators $a_{\lambda,l,\vb*{k}}$ is the annihilation operator of magnon at $l$ sub lattice of layer λ. 
In this model, the magnon Hamiltonian $H_{\text{m}_\lambda}(\vb*{k})$ takes the same form for the top and bottom layers, which is given by
\begin{equation}
    H_{\text{m}_\lambda}(\vb*{k}) = \begin{pmatrix}
        \Xi_\lambda(\vb*{k}) & \chi_\lambda(\vb*{k})\\
        \chi_\lambda^*(-\vb*{k}) & \Xi_\lambda^*(-\vb*{k})
        \end{pmatrix},\label{eq:magnon}
\end{equation}
with
\begin{align}
    &\Xi_\lambda(\vb*{k}) = \begin{pmatrix}
        \omega_{0,\vb*{k}}+\Delta_{\lambda,\vb*{k}} & J_1S\gamma(\vb*{k})\\
        J_1S\gamma^*(\vb*{k}) & \omega_{0,\vb*{k}}-\Delta_{\lambda,\vb*{k}}
        \end{pmatrix},\\
    &\chi_\lambda(\vb*{k}) = 0,
\end{align}
 where $\omega_{0,\vb*{k}}=3J_1S+6J_2S-2J_2S\sum_l\cos{(\vb*{k}\cdot\vb*{d}_l)}+2\Delta_zS$ and $\Delta_{\lambda,\vb*{k}}=2SD\sum_l\sin{(\vb*{k}\cdot\vb*{d}_l)}$.

The magnon Hamiltonian reveals that the DM interaction breaks the TRS of magnon Hamiltonian. Furthermore, in this model, the DM interaction plays a role analogous to the SOC in the Haldane model. Indeed, the DM interaction in this model is known to induce a topological magnon band structure analogous to that of the Haldane model\cite{Owerre2016AInsulator}.

The interaction between electrons and spins in Eq.~\eqref{eq:interaction} leads to the interaction between electrons and magnons as
\begin{align}
    \mathcal{H}_{\text{int}} = &-\sqrt{2S}J_{ex}\sum_{\vb*{k},\vb*{q}}\sum_l(c_{l,\downarrow,\vb*{k}}^\dagger c_{l,\uparrow,\vb*{q}}a_{b,l,\vb*{k}-\vb*{q},}+h.c.)\notag\\
   &-\sqrt{2S}J_{ex}\sum_{\vb*{k},\vb*{q}}\sum_l(-c_{l,\downarrow,\vb*{k}}^\dagger c_{l,\uparrow,\vb*{q}}a^\dagger_{t,l,\vb*{k}-\vb*{q}}+h.c.). \label{eq:int}
\end{align}
In this model, the top and bottom layers have an opposite spin direction. As a result, interactions $\mathcal{H}_{\text{int}0}$ and $\mathcal{H}_{\text{int}2}$ vanish, and the exchange field and $\Sigma_{cor}$ become zero. Furthermore, $\tilde{V}^{\gamma\gamma\delta}_{\vb*{k},\vb*{k}}=0$ and the tadpole term $\Sigma_{tad}$ vanishes in the present case.
The effect of electron-spin coupling is therefore described solely by the scattering term $\Sigma_{sca}$.
Furthermore, in this model, the electron Hamiltonian $\mathcal{H}_{\text{e}}$ is not renormalized by static effects ($\mathcal{H}_{\text{e}}=\mathcal{H}^\prime_{\text{e}}$) and total spin $S^z$ is conserved. Therefore, the electron Hamiltonian and self-energy can be decoupled into independent spin-up and spin-down sectors, which simplifies the self-energy calculation. The self-energy matrix becomes diagonal in the spin indices and can be evaluated using the expression in Eq.~\eqref{eq:self_energy_decoupled}. 
Consequently, the total self-energy $\Sigma^{\alpha\beta,\mu\nu}_{\vb*{k},\omega}=\Sigma^{\alpha\beta,\mu\nu}_{sca,\vb*{k},\omega}+\Sigma^{\alpha\beta,\mu\nu}_{cor,\vb*{k},\omega}+\Sigma^{\alpha\beta,\mu\nu}_{tad,\vb*{k},\omega}=\Sigma^{\alpha\beta,\mu\nu}_{sca,\vb*{k},\omega}$ satisfies
\begin{align}
    \Sigma^{\alpha\beta,\uparrow\downarrow}_{\vb*{k},\omega}=\Sigma^{\alpha\beta,\downarrow\uparrow}_{\vb*{k},\omega}=0.
\end{align}

\begin{figure}[tb]
    \centering
    \includegraphics[width=\linewidth]{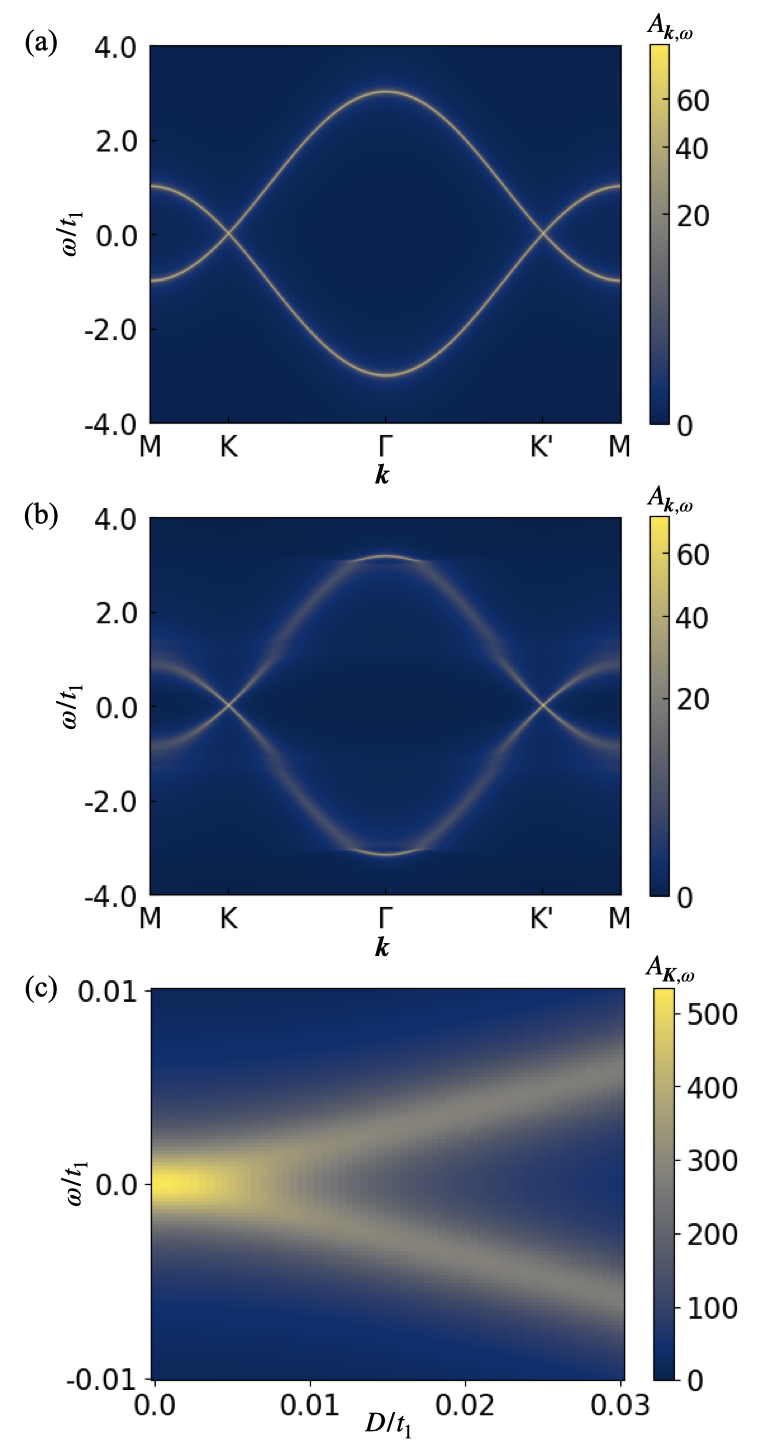}
    \caption{Spectral function of electrons. (a) A spectral function without self-energy along the high-symmetry path ($\eta/t_1=0.01$). (b) A spectral function with self-energy ($D/t_1=0.0$, $\eta/t_1=0.01$). (c) Spectral function $A_{\vb*{k},\omega}$ at the $K$ point as a function of DM interaction ($\eta/t_1=0.002$). We use following parameters: $t_2/t_1=0.0$, $J_1/t_1=0.05$, $J_2/t_1=0.01$, $J_{ex}/t_1=0.5$, $S=1$, $\Delta_z/t_1=0.005$ and $\varepsilon_F/t_1=0.0$} 
    \label{fig:bilayer_spectl}
\end{figure}

\subsection{Spectral function and effective Hamiltonian}
With the self-energy obtained in the above, we can compute the renormalized Green's function of the electron system,
\begin{equation}
    \mathcal{G}_{e,\vb*{k},\omega} = (\omega-\varepsilon_{\vb*{k}}-\Sigma_{\vb*{k},\omega}+i\eta)^{-1},
\end{equation}
and the spectral function,
\begin{equation}
    A_{\vb*{k},\omega} = -\frac{1}{\pi}\text{Im}\mathcal{G}_{e,\vb*{k},\omega}.
\end{equation}

We show the result for the spectral function in Fig.\ref{fig:bilayer_spectl}. 
Without the electron-magnon interaction, the spectral function, determined by $\mathcal{G}_{e,\vb*{k},\omega}^0$ has a Dirac cone structure at the $K$ point as shown in Fig.\ref{fig:bilayer_spectl} (a). When we consider the electron-magnon interaction, the spectral function is modified as shown in Fig.~\ref{fig:bilayer_spectl} (b). In particular, the spectral function around the $\Gamma$ point is modified and energy bands are shifted from their original positions in the band structure. This behavior can be understood by considering the properties of the self-energy.
Assuming that the magnon energy is significantly smaller than the electron energy, Eq.~\eqref{eq:self_energy_decoupled} indicates that the imaginary part of the self-energy, $\text{Im}\Sigma(\omega)$, becomes large when there exist electronic states at the energy $\omega$ satisfying $\omega = \varepsilon_{\vb*{q}}$ at some momentum $q$.
In particular, when the density of states (DOS) at $\omega$ is high, $\text{Im}\Sigma(\omega)$ is substantially enhanced, leading to a reduction in the electron lifetime. 

In contrast, the spectral function around the $K$ point is not significantly modified by the electron-magnon interaction since the DOS around the $K$ point is low. Therefore, the linear dispersion around the Dirac point at the Fermi energy 
remains clear even in the presence of the electron-magnon interaction.

Now, we focus on the effect of the DM interaction. The DM interaction breaks the effective TRS and induces topological magnons. We show that the DM interaction modifies electron bands through the electron-magnon interaction. We focus on the Dirac point at the $K$ point and evaluate the energy gap as a function of the DM interaction in Fig.\ref{fig:bilayer_spectl} (c).
The opening of an energy gap, driven by the DM interaction in the spin Hamiltonian, indicates a transfer of TRS-breaking and topology from the magnon band to the electronic system via the electron-magnon interaction.

Evaluating physical quantities such as the Chern number and Hall conductivity directly from the Green's function requires integration over both momentum and frequency \cite{Wang2010TopologicalInsulators}, which is computationally demanding. To simplify this process, we construct an effective Hamiltonian that captures the essential topological features of the electron system under the magnon interactions.
Here, we consider the effective Hamiltonian around the Dirac point since the Berry curvature becomes large around the Dirac point.
By substituting the energy of the Dirac point with the frequency of self-energy, we approximate the self-energy $\Sigma_{\vb*{k},\omega}^{\alpha\beta,\mu\mu}$as $\Sigma_{\vb*{k},\varepsilon_{D,\mu}}^{\alpha\beta,\mu\mu}$ where $\varepsilon_{D,\mu}$ is the energy of the Dirac point. Then, by using the approximated self-energy, we obtain the effective Hamiltonian as
\begin{equation}
    H_{\text{eff},\vb*{k},\mu} = H_{\text{e},\vb*{k},\mu}^{\prime}+T_{\vb*{k},\mu}\frac{\Sigma^{\mu\mu}_{\vb*{k},\varepsilon_{D,\mu}}+(\Sigma^{\mu\mu}_{\vb*{k},\varepsilon_{D,\mu}})^\dagger}{2}T_{\vb*{k},\mu}^\dagger\label{eq:effective_hamiltonian}
\end{equation}
where $\Sigma^{\mu\mu}_{\vb*{k},\varepsilon_{D,\mu}}$ is the self-energy matrix with spin $\mu$ and $T_{\vb*{k},\mu}$ is the unitary matrix which diagonalizes the Hamiltonian $H_{\text{e},\vb*{k},\mu}^{\prime}$.

\begin{figure}[tbp]
    \centering
    \includegraphics[width=\linewidth]{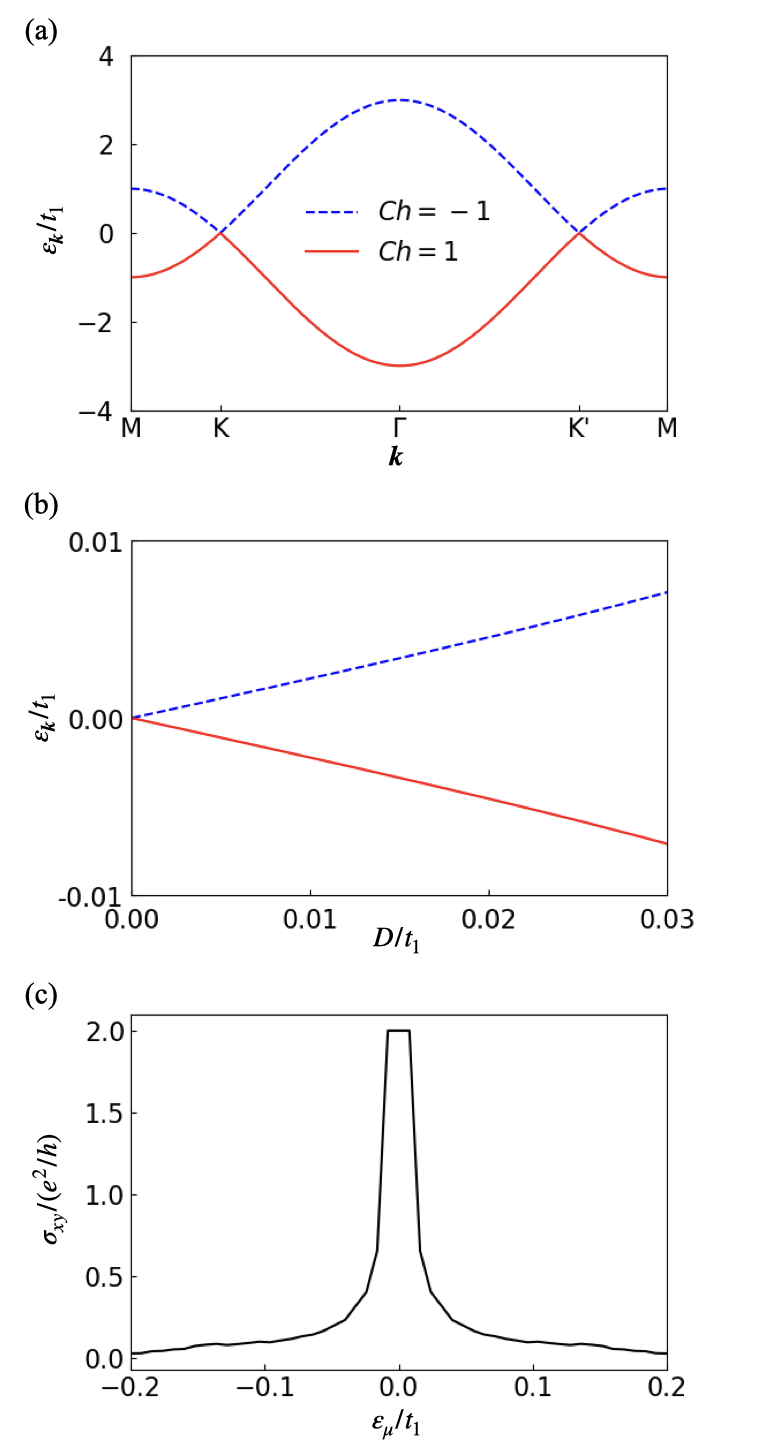}
    \caption{Band energy and the Hall conductivity of the effective Hamiltonian of a multi-layer system. (a) Energy band of the effective Hamiltonian ($\varepsilon_F=0.0$). (b) The energy gap at the $K$ point of the effective Hamiltonian as a function of DM interaction. (c) Hall conductivity $\sigma_{xy}$ as a function of the $\varepsilon_F$ at the zero temperature. We use following parameters: $t_2/t_1=0.0$, $J_1/t_1=0.05$, $J_2/t_1=0.01$, $D/t_1=0.02$, $J_{ex}/t_1=0.5$, $S=1$, $\Delta_z/t_1=0.005$ and $\eta/t_1=0.02$.} 
    \label{fig:bilayer}
\end{figure}

From this effective Hamiltonian, we can obtain the Berry curvature, Chern number, and Hall conductivity $\sigma_{xy}$. 
Figure \ref{fig:bilayer} shows the resulting energy band, the energy gap around the $K$ point, and the Hall conductivity derived from the effective Hamiltonian. A comparison of the calculated bands (Fig.\ref{fig:bilayer}(a)) with the spectral function (Fig.\ref{fig:bilayer_spectl}(a)) reveals that the effective Hamiltonian does not capture the shift of energy bands around the $\Gamma$ point. However, the energy band of the effective Hamiltonian and the spectral function are in qualitative agreement around the Dirac point, allowing us to describe the topological gap opening.
Furthermore, the energy band of the effective Hamiltonian has non-zero Chern numbers ($Ch=\pm 1$), as indicated in Fig.\ref{fig:bilayer} (a). Therefore, the effective Hamiltonian describes a Chern insulator.

The energy gap at the $K$ point of the effective Hamiltonian is shown in Fig.\ref{fig:bilayer} (b). The energy gap is proportional to the DM interaction, indicating the transfer of magnon topology to the electron band. 
We can see that the energy gap of the effective Hamiltonian in Fig.\ref{fig:bilayer} (b) is consistent with the energy gap of the spectral function in Fig.\ref{fig:bilayer_spectl} (c).

Now, we evaluate the Hall conductivity $\sigma_{xy}$ of the effective Hamiltonian by using the Kubo formula
\begin{equation}
    \sigma_{xy} = \frac{e^2}{hN}\sum_{\vb*{k}}\sum_{\alpha} f({\varepsilon_{\mathrm{eff},\vb*{k},\alpha}})\Omega_{\vb*{k},\alpha},
\end{equation}
where $\Omega_{\vb*{k}}$ is the Berry curvature and $\varepsilon_{\mathrm{eff},\vb*{k}}$ is an energy of the effective Hamiltonian.
The Hall conductivity as a function of the chemical potential $\varepsilon_F$ is shown in Fig.~\ref{fig:bilayer}(c). A quantized plateau develops around $\varepsilon_F=0$, since the Dirac point is at the Fermi energy with $\varepsilon_F=0$. Also, a quantized value is $2e^2/h$, since the up-spin and down-spin bands are degenerate and have the same Chern number ($\pm1$).
This indicates that the electron system becomes a quantum Hall insulator, even though it is topologically trivial without the magnon system. In other words, the topology of the magnon system is imprinted on the electron system through the electron-magnon interaction. 
Here, we evaluate the Hall conductivity using the effective Hamiltonian, neglecting factors such as the electron lifetime. Therefore, the Hall conductivity obtained here is an approximation of that derived from the Green's function. Since the spectral function exhibits a gap around the Fermi level which can be well captured by the effective Hamiltonian, it is expected that the full Green's function calculation also yields quantized Hall conductivity as far as the Fermi level lies within this gap.

While we focus on the multilayer system with zero net exchange field in this section, our analysis is also applicable to other systems, such as bilayer systems or magnetic metals. In those systems, the electron-spin interaction typically induces a finite net exchange field. The strength of the exchange field is tunable by temperature, as it is proportional to the expectation value of the local spin, $\Braket{S^z}$. As we demonstrate for a representative model in Appendix~\ref{sec:appendix_model}, such temperature dependence provides a mechanism to modify the energy of the topological gap, potentially driving the topological phase transition into a quantum Hall insulator (QHI) phase.

\section{Quantum Spin Hall insulator and $Z_2$ topological phase induced by magnon} \label{sec:QSHI}
In this section, we consider a time-reversal symmetric system. In particular, we consider a topologically trivial metal with TRS, sandwiched by two spin systems that are time-reversal (TR) partners of each other. While each individual spin layer breaks TRS, the total Hamiltonian remains TR-invariant.

\subsection{Quantum spin Hall insulator}

\begin{figure}[htb]
    \centering
    \includegraphics[width=\linewidth]{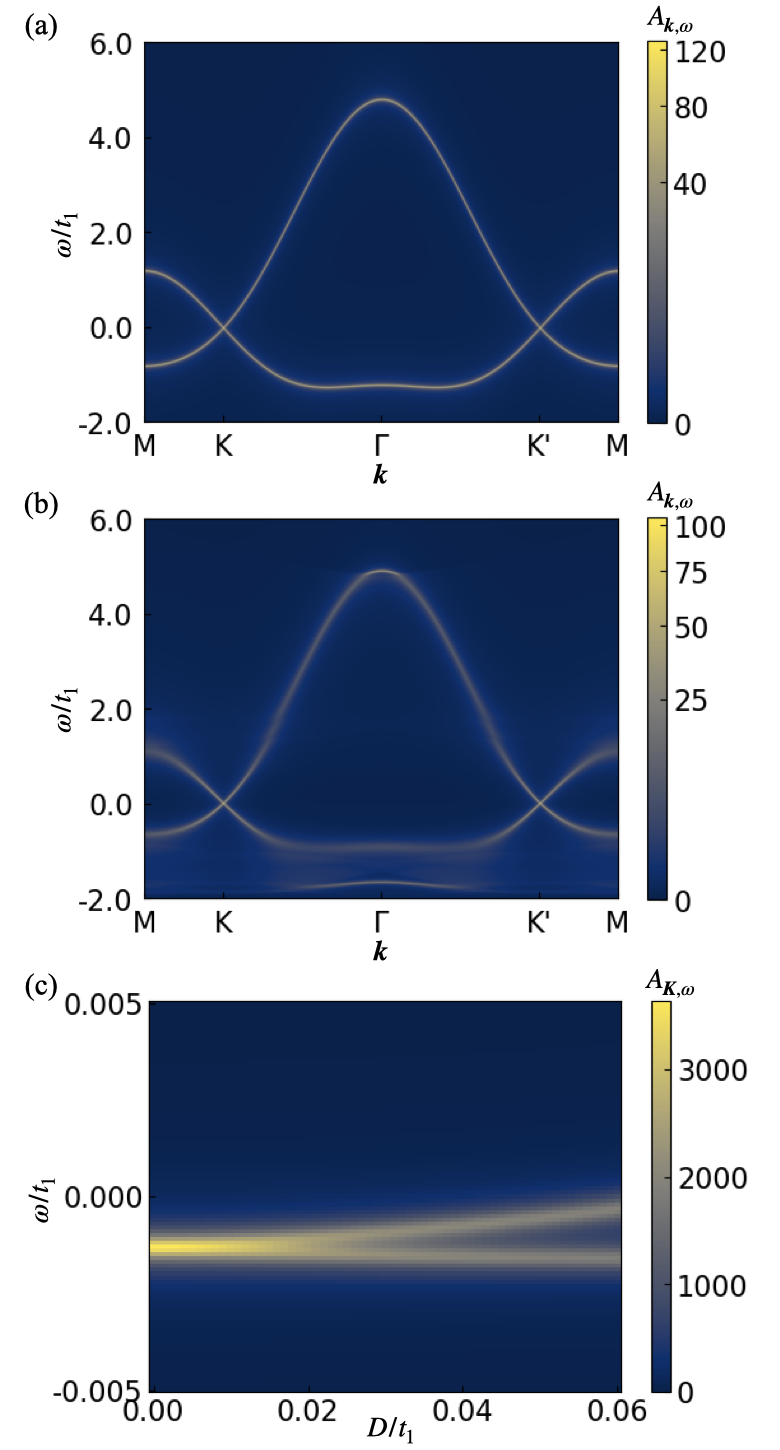}
    \caption{Spectral function of electrons. (a) A spectral function without self-energy along the high-symmetry path ($\eta/t_1=0.01$). (b) A spectral function with self-energy ($D/t_1=0.0$, $\eta/t_1=0.01$). (c) Spectral function $A_{\vb*{k},\omega}$ at the $K$ point as a function of DM interaction ($\eta/t_1=0.0003$). We use following parameters: $t_2/t_1=-0.2$, $J_1/t_1=0.1$, $J_2/t_1=0.02$, $D/t_1=0.04$, $J_{ex}/t_1=1.0$, $S=1$, $\varepsilon_F/t_1=-0.57$, $\Delta_z/t_1=0.01$ and $k_BT=0.0$.} 
    \label{fig:QSHI_spectl}
\end{figure}

We consider a three-layer honeycomb model where an electron system is sandwiched by spin systems in the top and bottom layers that are time-reversal pairs with each other. 
Similarly to the previous section, 
the total Hamiltonian is given by Eq.~\eqref{eq:total_hamiltonian} with the electron Hamiltonian given by Eq.~\eqref{eq:electron}, the spin Hamiltonian of bottom layer is given by Eq.~\eqref{eq:spin_model} and the interaction Hamiltonian given by Eq.~\eqref{eq:interaction}. 
The spin Hamiltonian of top layer is now a time-reversal pair of the bottom layer which has the same sign for the DM interaction as $\xi_{t,ij}=2/(\sqrt{3}a^2)\vb*{b}_i\times\vb*{b}_j=\xi_{b,ij}$.

In this model, the spin systems in the top and bottom layers break effective TRS, but the total Hamiltonian has a TRS because the spin systems in the top and bottom layers are time-reversal partners. 

By using the Holstein-Primakoff transformation, the magnon Hamiltonian for the bottom layer $H_{\text{m}_b}$ is again given by Eq.~\eqref{eq:magnon}, and the interactions between electrons and magnons are given by Eq.~\eqref{eq:int}. 
Since the magnon Hamiltonian of the top layer is modified to be a time-reversal pair of the bottom layer, the magnon Hamiltonian of the top layer $H_{m_t}(\vb*{k})$ is given by the form of Eq.~\eqref{eq:magnon} with setting $\Delta_{t,\vb*{k}}=-2SD\sum_l \sin(\vb*{k}\cdot\vb*{d}_l)$.

In this system, the electron system is a topologically trivial metal without an interaction with the magnons. With a coupling to magnon systems in the bottom and top layers having opposite Chern numbers and opposite spins, the combined magnon subsystem constitutes an analog of the Kane-Mele model and becomes topologically non-trivial. 
Namely, the nontrivial magnon topology is transferred to the electron system via the electron-magnon interaction, inducing a quantum spin Hall insulator (QSHI) phase.

The electronic spectral function is presented in Fig.~\ref{fig:QSHI_spectl}. In the absence of the electron-magnon interaction, the spectrum is given by the unperturbed Green's function and exhibits a Dirac cone at the $K$ point [Fig.~\ref{fig:QSHI_spectl}(a)]. Upon introducing the electron-magnon interaction [Fig.~\ref{fig:QSHI_spectl}(b)], the spectral function is significantly modified. 

Similarly to the quantum Hall Insulator (QHI) case (cf. Fig.~\ref{fig:bilayer_spectl}), the lifetime of electrons is reduced due to the large DOS. Furthermore, under these conditions, the real part of the self-energy also increases. In particular, the off-diagonal part of the self-energy becomes large. Thus, the pole of the Green's function splits into two. Consequently, the spectral weight near the $\Gamma$ point, where the DOS is large, is pushed out from its original energy region, forming a new subband around $\omega/t_1\sim-2.0$. However, the Dirac cone at the $K$ point persists since the DOS is small around $K$ point. Furthermore, the energy of the Dirac point is shifted by the electron-magnon interaction. This shift is due to the real part of the self-energy, which modifies the energy of the electron.
Focusing on the Dirac point, the energy gap opens due to the DM interaction as shown in Fig.\ref{fig:QSHI_spectl} (c). 

\begin{figure}[htb]
    \centering
    \includegraphics[width=\linewidth]{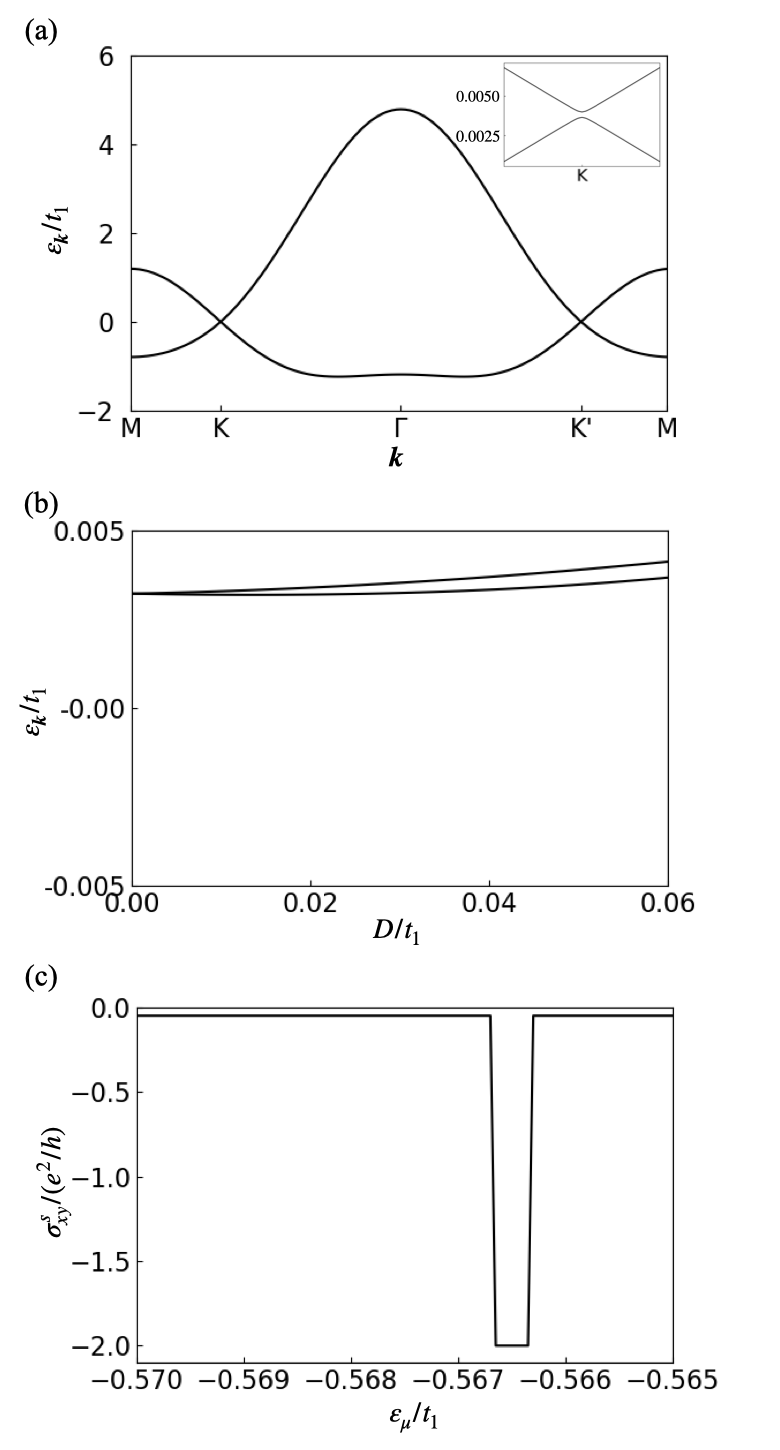}
    \caption{Band energy and the Hall conductivity of the effective Hamiltonian of QSHI. (a) The energy band of the effective Hamiltonian ($\varepsilon_F=-0.57$). The inset is an enlarged view of the gap structure around the $K$ point. (b) The energy gap at the $K$ point of the effective Hamiltonian as a function of DM interaction.
    (c) The spin Hall conductivity $\sigma^{S^z}_{xy}$ as a function of $\varepsilon_F$. We use following parameters: $t_2/t_1=-0.2$, $J_1/t_1=0.1$, $J_2/t_1=0.02$, $D/t_1=0.04$, $J_{ex}/t_1=0.5$, $S=1$, $\Delta_z/t_1=0.01$, $k_BT=0.0$ and $\eta/t_1=0.02$.} 
    \label{fig:QSHI}
\end{figure}

To investigate the topology of the electron Hamiltonian, we construct an effective Hamiltonian around the Dirac point in the same way as in the previous section with Eq.~\eqref{eq:effective_hamiltonian}. 
The energy band of the effective Hamiltonian is shown in Fig.~\ref{fig:QSHI} (a). The energy band has a Dirac cone structure at the $K$ point and the energy gap opens due to the DM interaction. In this model, the $z$ component of spin $S^z$ is conserved and the up-spin and down-spin bands are degenerate having opposite Chern numbers. The energy gap increases monotonically with the strength of the DM interaction, as shown in Fig.~\ref{fig:QSHI}(b). 
The effective model qualitatively reproduces the gap structure in the full spectral function, although the self-energy causes an energy shift of the gap position in the full spectral function.

Due to the degenerate spin bands with opposite Chern numbers, the net charge Hall conductivity vanishes. 
In contrast, a non-zero spin Hall conductivity $\sigma_{xy}^z$ can be defined as $\sigma_{xy}^z=\sigma_{xy,\uparrow}-\sigma_{xy,\downarrow}$, where $\sigma_{xy,\uparrow}$ ($\sigma_{xy,\downarrow}$) is the Hall conductivity of the up-spin (down-spin) band. As shown in Fig.~\ref{fig:QSHI}(c), the spin Hall conductivity $\sigma_{xy}^z$ exhibits a quantized plateau when the Fermi energy lies within the topological gap. This demonstrates the realization of a QSHI, where the non-trivial topology of the magnon system is imprinted onto the electron system.

\subsection{$Z_2$ topological phases}
In this subsection, we consider the system where spin is not conserved, where the magnons induce an effective SOC to the electron system. Consequently, the system's topology can no longer be described by a spin-resolved Chern number; instead, it is characterized by the $Z_2$ topological invariant.

We consider the Hamiltonian in Eq.~\eqref{eq:total_hamiltonian}
where we modify the spin Hamiltonian to induce the broken inversion symmetry and the SOC. 
Specifically, we add a staggered spin-axis anisotropy to the spin Hamiltonian for the bottom layer, $\mathcal{H}_{\text{S}_\lambda}$,  which is now written as
\begin{align}
    \mathcal{H}_{\text{S}_\lambda}&=\sum_{\langle i,j \rangle}-J_1\vb*{S}_i\cdot \vb*{S}_j\notag\\
    &+\sum_{\langle\langle i,j \rangle\rangle}[-J_2\vb*{S}_i\cdot \vb*{S}_j-D\vb*{\xi}_{\lambda,ij}\cdot(\vb*{S}_i\times \vb*{S}_j)]\notag\\
    &-\sum_{i} [\Delta_{z}^i(\sin\theta S_i^{x}+\cos\theta S_i^{z})^2],
\end{align}
where $\Delta_{z}^i$ is the staggered spin-axis anisotropy with $\Delta_{z}^i = \Delta_z^A$ for the sublattice A and $\Delta_{z}^i = \Delta_z^B$ for the sublattice B. The factor $\vb*{\xi}_{\lambda,ij}=2/(\sqrt{3}a^2)\vb*{b}_i\times\vb*{b}_j$.  Here, we assume that the ground state of the spin system is ferromagnetic, where $\vb*{S}_i=S(\sin\theta,0,\cos\theta)$ for the bottom layer and $\vb*{S}_i=-S(\sin\theta,0,\cos\theta)$ for the top layer, respectively. The form of this Hamiltonian indicates that the spin component $S^z$ is not conserved, and the staggered anisotropy term explicitly breaks the inversion symmetry.
By rotating the spin quantization axis, the spin Hamiltonian $\mathcal{H}_{\text{S}_\lambda}$ can be expressed in terms of the transformed spin operators $\tilde{\vb*{S}}$ as
\begin{align}
    \mathcal{H}_{\text{S}_\lambda} = &-J_1\sum_{\langle i,j \rangle}\tilde{\vb*{S}}_i\cdot \tilde{\vb*{S}}_j+ \sum_{\langle\langle i,j \rangle\rangle}[-J_2\tilde{\vb*{S}}_i\cdot \tilde{\vb*{S}}_j\notag\\
    &-D\tilde{\xi}^z_{\lambda,ij}\{\cos\theta(\tilde{\vb*{S}}_i\times\tilde{\vb*{S}}_j)_z-\sin\theta(\tilde{\vb*{S}}_i\times\tilde{\vb*{S}}_j)_x\}]\notag\\
    &-\sum_i \Delta^i_z(\tilde{S}^{z}_i)^2,
\end{align}
where $\tilde{\vb*{\xi}}_{b,ij}=\vb*{\xi}_{\lambda,ij}$ and $\tilde{\vb*{\xi}}_{t,ij}=-\vb*{\xi}_{\lambda,ij}$.
Applying the Holstein-Primakoff transformation, we obtain the magnon Hamiltonian for the bottom layer. This Hamiltonian takes the form in Eq.~\eqref{eq:magnon}, with the matrix elements given by
\begin{align}
    &\Xi_\lambda(\vb*{k}) = \begin{pmatrix}
        \omega_{A,\vb*{k}}+\Delta_{\lambda,\vb*{k}}\cos\theta& J_1S\gamma(\vb*{k})\\
        J_1S\gamma^*(\vb*{k}) & \omega_{B,\vb*{k}}-\Delta_{\lambda,\vb*{k}}\cos\theta
        \end{pmatrix},\\
    &\chi_\lambda(\vb*{k}) = 0,
\end{align}
where $\omega_{A,\vb*{k}}=3J_1S+6J_2S-2J_2S\sum_l\cos{(\vb*{k}\cdot\vb*{d}_l)}+2\Delta_z^AS$, $\omega_{B,\vb*{k}}=3J_1S+6J_2S-2J_2S\sum_l\cos{(\vb*{k}\cdot\vb*{d}_l)}+2\Delta_z^BS$, $\Delta_{b,\vb*{k}}=2SD\sum_l\sin{(\vb*{k}\cdot\vb*{d}_l)}$ and $\Delta_{t,\vb*{k}}=-2SD\sum_l\sin{(\vb*{k}\cdot\vb*{d}_l)}$. The magnon Hamiltonian for the top layer is given by the time-reversal pair of the above Hamiltonian.
The interaction between electron and magnon is given by
\begin{align}
    \mathcal{H}_{\text{int}} = &-\sqrt{\frac{S}{2}}J_{ex}\sum_{\vb*{k},\vb*{q}}\sum_{l}[(\cos\theta\sigma^x_{\mu\nu}-i\sigma^y_{\mu\nu}-\sin\theta\sigma^z_{\mu\nu})\notag\\
    &\times c_{l,\mu,\vb*{k}}^\dagger c_{l,\nu,\vb*{q}}a_{b,l,\vb*{k}-\vb*{q}}+h.c.]\notag\\
     &-\sqrt{\frac{S}{2}}J_{ex}\sum_{\vb*{k},\vb*{q}}\sum_{l}[(-\cos\theta\sigma^x_{\mu\nu}-i\sigma^y_{\mu\nu}+\sin\theta\sigma^z_{\mu\nu})\notag\\
    &\times c_{l,\mu,\vb*{k}}^\dagger c_{l,\nu,\vb*{q}}a_{t,l,\vb*{k}-\vb*{q}}+h.c.]
    \label{eq:int_Z2}
\end{align}
For a canted spin configuration ($\theta\neq0$), this interaction couples  electronic states with up and down spins. Following the method in Sec~\ref{sec:QH}, we construct an effective Hamiltonian to investigate the resulting topology of the electron system. When $\theta\neq0$, though the relevant interaction vertex $\tilde{V}^{\gamma\gamma\delta}_{\vb*{k},\vb*{k}}$ is nonzero, the self-energy $\Sigma_{tad}$ becomes zero. This is because $\tilde{V}$ satisfies $\tilde{V}^{\gamma\gamma\delta,\uparrow\uparrow}_{\vb*{k},\vb*{k}}=-\tilde{V}^{\gamma\gamma\delta,\downarrow\downarrow}_{\vb*{k},\vb*{k}}$ and $\varepsilon_{\mu,\vb*{k}}$ satisfies $\varepsilon_{\uparrow,\vb*{q},\gamma}=\varepsilon_{\downarrow,\vb*{q},\gamma}$, $\Sigma_{tad}=0$.

\begin{figure}[htbp]
    \centering
    \includegraphics[width=\linewidth]{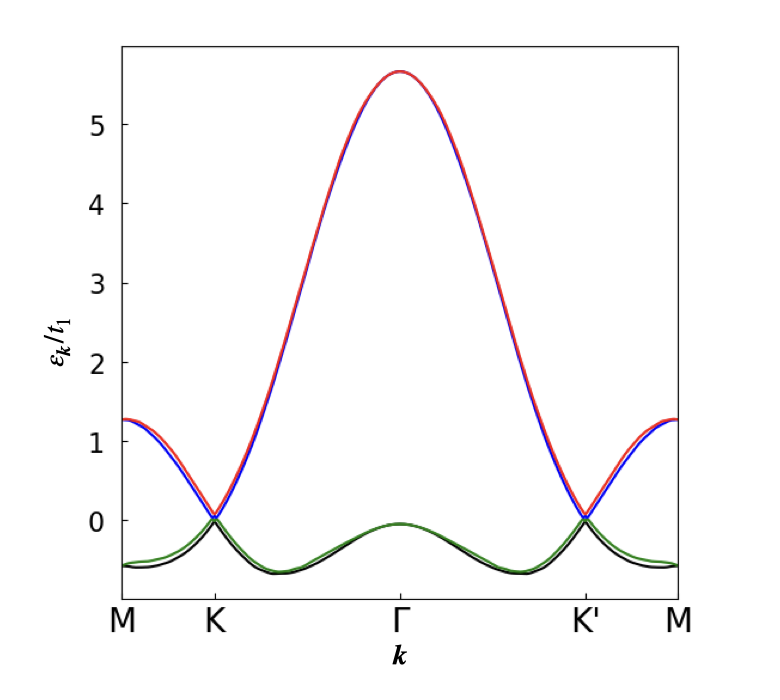}
    \caption{The band structure of effective Hamiltonian. We use following parameters: $t_2/t_1=-0.3$, $J_1/t=0.1$, $J_2/t=0.02$, $D/t_1=0.05$, $J_{ex}/t=1.5$, $S=1$, $\Delta_z^A/t=0.15$, $\Delta_z^B/t=0.05$, $\varepsilon_F/t=-0.4$, $\theta=0.5$ and $\eta/t_1=0.02$.} 
    \label{fig:Z2_band}
\end{figure}

Figure~\ref{fig:Z2_band} shows the band structure of the effective Hamiltonian. Due to the broken inversion symmetry and the effective SOC, the spin degeneracy is lifted throughout the Brillouin zone, except at the time-reversal invariant momenta (TRIM)—the $\Gamma$ and $M$ points—where Kramers degeneracy is preserved.

\begin{figure}[htbp]
    \centering
    \includegraphics[width=\linewidth]{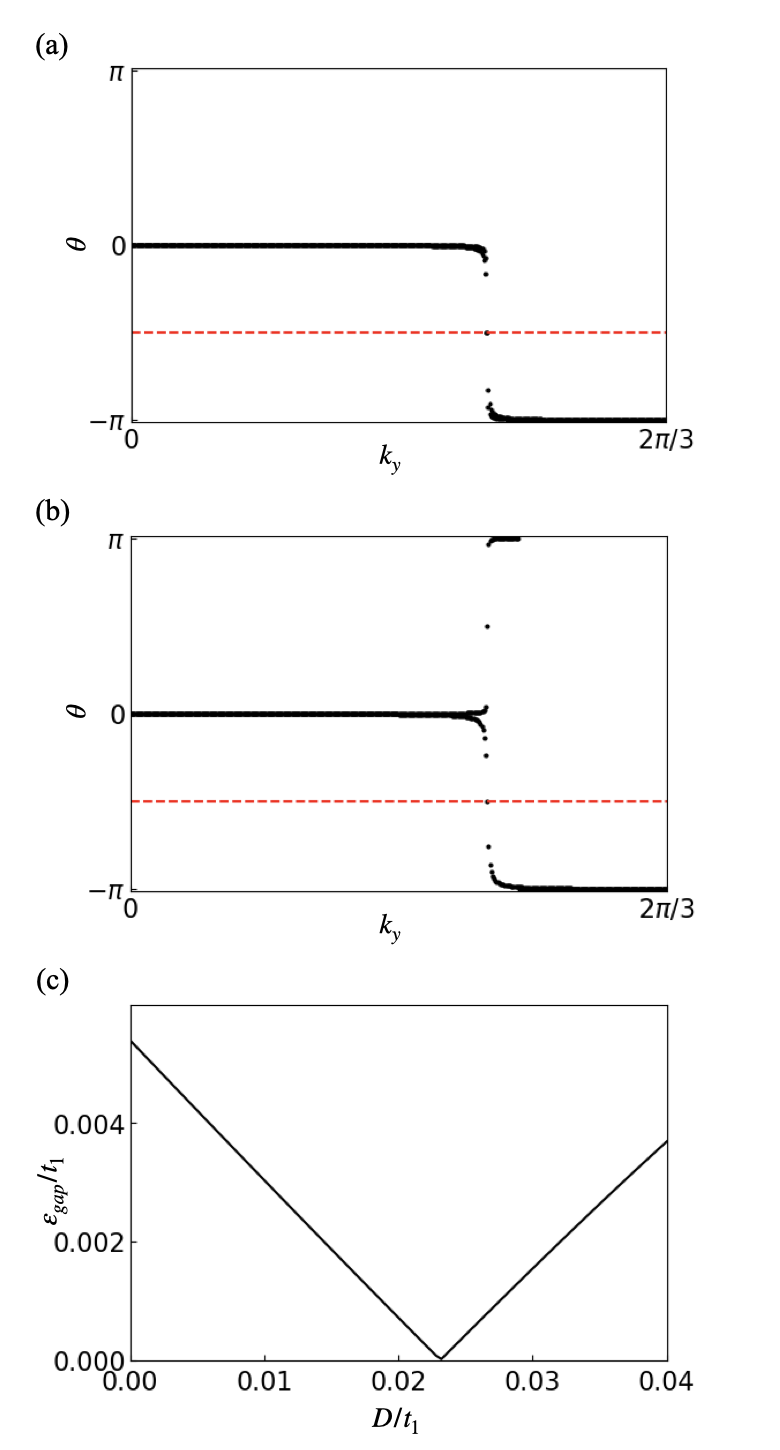}
    \caption{$Z_2$ index and band gap of the effective Hamiltonian. (a) The Wannier center $\theta(k_y)$ for small $D$ ($D/t_1=0.01$). A red dashed line indicates reference line. (b) The Wannier center $\theta(k_y)$ for large $D$ ($D/t_1=0.03$). (c) The energy gap as a function of the DM interaction.
    We use following parameters:$t_2/t_1=-0.3$, $J_1/t=0.1$, $J_2/t=0.02$, $J_{ex}/t=0.8$, $S=1$, $\Delta_z^A/t=0.017$, $\Delta_z^B/t=0.003$, $\varepsilon_F/t=-0.745$, $\theta=0.5$ and $\eta/t_1=0.02$.} 
    \label{fig:Wilson}
\end{figure}

To characterize the system's topology, we calculate the $Z_2$ invariant by tracking the evolution of the Wannier function centers~\cite{Yu2011EquivalentConnection} (see Appendix\ref{appendix:Z2}). 
Let us consider the Wilson loop for occupied bands,
\begin{equation}
    W(k_y)=\mathcal{P}\exp{\bigg[\int_{C_{k_y}} -iA(\vb*{k})dk\bigg]}
\end{equation}
where $\mathcal{P}$ is a path-ordering operator, $A(\vb*{k})$ is a Berry connection matrix for occupied bands, and $C_{k_y}$ denotes a loop in the momentum space which goes across the Brillouin zone with fixed $k_y$. The Wilson loop is a $N_{occ}\times N_{occ}$ matrix ($N_{occ}$: the number of occupied states) and the Wannier function centers for the states of fixed $k_y$ are characterized by the phase $\theta(k_y)$ of eigenvalues of the Wilson loop $W(k_y)$.
The evolution of the Wannier function centers and the energy gap at $K$ point are represented in Fig.~\ref{fig:Wilson}. For a small DM interaction($D/t_1=0.01$), the evolution of the Wannier function centers $\theta (k_y)$ crosses an arbitrary reference line an even number of times (see Fig.~\ref{fig:Wilson}(a)), indicating a trivial insulator. In contrast, for a large DM interaction ($D/t_1=0.03$), $\theta (k_y)$ crosses an arbitrary reference line an odd number of times (see Fig. \ref{fig:Wilson}(b)), indicating a $Z_2$ topological phase. 
Figure \ref{fig:Wilson}(c) shows the energy gap at $K$ point. In this model, the gap opens even though the DM interaction is zero. This is because the spin axis anisotropy breaks the inversion symmetry and opens a gap at the $K$ point.
The gap magnitude decreases as the DM interaction increases, and the gap closes near $D/t_1 \sim 0.02$, which indicates a topological phase transition. For the present parameter sets in Fig.~\ref{fig:Wilson}, the Dirac point is slightly displaced away from the Fermi surface, and a metallic state appears in the vicinity of $D/t_1\sim0.02$, while the system becomes an insulator away from this parameter region. Namely, as the DM interaction  increases, the system undergoes a phase transition from a trivial insulator to an intermediate metallic state and then to a $Z_2$ topological insulator. Since $S^z$ is not conserved in this system, the spin current is not a conserved quantity and the spin Hall conductivity is not a well-defined quantity and does not show a quantized behavior. However, we can observe the spin current from the helical edge states.

\section{Discussion}\label{sec:discussion}
\subsection{Scope of the theory}
In this study, we show that the topology of magnons can be imprinted on the electron systems through the electron-magnon interactions. Our analysis is applicable to magnetic metals and metal/magnet heterostructures. The mechanism for topological electronic states proposed in this study is particularly useful for systems in which the spin system breaks the effective TRS even if the electron system itself is time-reversal symmetric and is topologically trivial. For simplicity, we considered spin systems involving ferromagnetic order and DM interactions. Recently, it has been theoretically proposed that the Kitaev interaction may break the TRS of the spin system and give rise to a topological magnon, and the present theory is applicable to those systems as well.

Furthermore, the present mechanism is also applicable to the case where the ground state of the spin system is not ferromagnetic. 
When the ground state is antiferromagnetically ordered, the staggered exchange field is induced and the inversion symmetry is broken, which may open a topologically trivial gap. Therefore, in such cases, it is necessary to consider a system with a stacking structure that preserves the inversion symmetry.

\subsection{Candidate materials}
The present mechanism for topological electronic states can be applied to magnetic metals and metal/magnet heterostructures. In particular, kagome and honeycomb materials are promising candidates since these structures can host topological magnons.
As an example of a metallic kagome-lattice ferromagnet, \ce{Fe_3Sn_2} has been studied extensively~\cite{Giefers2006HighSolution,Hou2017ObservationAnisotropy,Ye2018MassiveMetal,Yin2018GiantMagnet,Zhang2024SpinMetal}. It exhibits an orbital magnetic moment approximately five times larger than that of pure iron~\cite{Zhang2024SpinMetal} and an anomalous Hall effect has been observed~\cite{Ye2018MassiveMetal}. Furthermore, the band gap depends on the magnetization direction and cannot be explained by the simple Kane-Mele model~\cite{Yin2018GiantMagnet}. Given that the DM interaction in kagome lattices can give rise to topological magnons, it is plausible that the observed anomalous Hall effect is related to the presence of such magnons.

One of the other candidate materials is a heterostructure of graphene and \ce{CrI_3}~\cite{Zhang2018StrongHeterostructures,Farooq2019SwitchableHeterostructures,Holmes2020ExchangeHeterostructures,Tseng2022Gate-TunableInsulators}. In the honeycomb ferromagnet \ce{CrI_3} monolayer, a gap opening in the magnon spectrum at the $K$ point is observed~\cite{Chen2018TopologicalCrI3}, and theoretical studies suggest topological magnon~\cite{Costa2020TopologicalDescription,Brehm2024TopologicalFerromagnets}. 
In Ref.~\cite{Zhang2018StrongHeterostructures,Holmes2020ExchangeHeterostructures}, it is proposed that broken mirror symmetry induces SOC and induces an anomalous Hall effect in the graphene/\ce{CrI_3} heterostructure.
In our mechanism, however, even when mirror symmetry is preserved, the anomalous Hall effect can still emerge.  Here, we consider the trilayer structure \ce{CrI_3}/graphene/\ce{CrI_3} as demonstrated in Sec.~\ref{sec:QH}.
In our study, the topological gap becomes approximately $0.05t_1$ where $J_{ex}/t_1=0.5$ and scales with $J_{ex}^2$.
The effective magnetic field $J_{ex}S$ in graphene/\ce{CrI_3} is estimated to be $120$ meV~\cite{Zhang2018StrongHeterostructures,Farooq2019SwitchableHeterostructures,Holmes2020ExchangeHeterostructures} which can induce a gap of approximately $0.1$ to several meV to the electronic bands.

In addition to \ce{CrI_3}, \ce{MnPS_3} is a candidate material for topological magnons. \ce{MnPS_3} is a van der Waals material with a honeycomb lattice structure and exhibits antiferromagnetic order~\cite{Kurosawa1983NeutronFePS3}. Theoretical studies suggest that \ce{MnPS_3} is a candidate material for topological magnons~\cite{Zyuzin2016MagnonAntiferromagnets,Cheng2016SpinAntiferromagnets} and the magnon Nernst effect is observed~\cite{Shiomi2017ExperimentalMnPS3}. Since \ce{MnPS_3} is antiferromagnetic, the interaction between the electron and magnon breaks the inversion symmetry. Therefore, we consider the trilayer structure \ce{MnPS_3}/graphene/\ce{MnPS_3} to preserve the inversion symmetry and cancel out the staggered exchange field.

\subsection{Potential of interactions between quasiparticles.}
In this paper, we consider a model of topological magnons that can be described by a quadratic magnon Hamiltonian. However, recently other mechanisms for realizing topological magnon have been proposed theoretically. For example, topological magnons emerge due to the interaction between magnons~\cite{Mook2021Interaction-StabilizedFerromagnets,Sun2023InteractingMagnons} and the magnon-phonon interaction~\cite{Park2020ThermalInteraction,Go2019TopologicalNumbers,Go2024TopologicalTransition}. Our framework can be extended to systems where such topology arises from interactions between quasiparticles. Candidate materials where such interaction-driven topological magnons have been proposed include the kagome ferromagnet \ce{Cu(1,3-bdc)} and Kamiokite \ce{Fe2Mo3O8} \cite{Mook2021Interaction-StabilizedFerromagnets,Park2020ThermalInteraction}. Therefore, multilayer systems of metals and these magnets may also realize quantum Hall insulators.

\acknowledgements
We thank fruitful discussions with Hosho Kastura.
T.M. was supported by 
JSPS KAKENHI Grant 23K25816, 23K17665, and 24H02231.
K.F. was supported by JSPS KAKENHI Grant 24KJ0730 and the Forefront Physics and Mathematics program to drive transformation (FoPM).

\appendix

\section{Derivation of Formulation}\label{sec:appendix_derivation}
In this section, we show the derivation of Equations \eqref{General_Sigma_cor}, \eqref{General_Sigma_sca} and \eqref{General_Sigma_tad}.
The self-energy $\Sigma_{cor}$ is given by
\begin{align}
    \Sigma^{\alpha\beta}_{cor,\vb*{k},i\varepsilon_n} = \frac{k_BT}{N}\sum_{\vb*{q}}\sum_{\gamma}\sum_m \tilde{W}_{\vb*{k},\vb*{k},\vb*{q}}^{\alpha\beta\gamma\delta}\mathcal{G}^0_{m,\vb*{q},\gamma\delta}(i\omega_m),
\end{align}
where $\mathcal{G}^0_m(i\omega_m)$ is the Matsubara Green's function of a magnon with $\omega_m=2\pi im$. The Matsubara Green's function of a magnon in the imaginary-time formalism $\mathcal{G}^0_{m,\vb*{k},\alpha\beta}(\tau)$ is defined as
\begin{align}
    \mathcal{G}^0_{m,\vb*{k},\alpha\beta}(\tau)=&-\langle T_\tau \psi_{m,\vb*{k},\alpha}(\tau)\psi^\dagger_{m,\vb*{k},\beta}\rangle\notag\\
    =&-e^{-\sigma_{3,\alpha\alpha}\omega_{\vb*{k},\alpha}\tau}\sigma_{3,\alpha\beta}\notag\\
    &\times[\theta(\tau)n_B(-\sigma_{3,\alpha,\alpha}\omega_{{\vb*{k},\alpha}})\notag\\
    &-\theta(-\tau)(n_B(\sigma_{3,\alpha,\alpha}\omega_{\vb*{k},\alpha}))].
\end{align}
Fourier transforming $\mathcal{G}^0_m$ to the Matsubara frequency $\omega_n=2\pi in$, we obtain
\begin{align}
    \mathcal{G}^0_{m,\vb*{k},\alpha\beta}(i\omega_n) = \frac{\sigma_{3,\alpha\beta}}{i\omega_n-\sigma_{3,\alpha\alpha}\omega_{\vb*{k},\alpha}}.
\end{align}
Then, we can write $\Sigma_{cor}$ as
\begin{align}
    \Sigma^{\alpha\beta}_{cor,\vb*{k},i\varepsilon_n} &= \frac{k_BT}{N}\sum_{\vb*{q}}\sum_{\gamma}\sum_m \tilde{W}_{\vb*{k},\vb*{k},\vb*{q}}^{\alpha\beta\gamma\delta}\frac{\sigma_{3,\gamma\delta}}{i\omega_m-\sigma_{3,\gamma\gamma}\omega_{\vb*{q},\gamma}}\notag\\
    &=\frac{1}{N}\sum_{\vb*{q}}\sum_{\gamma} \tilde{W}_{\vb*{k},\vb*{k},\vb*{q}}^{\alpha\beta\gamma\gamma}\sigma_{3,\gamma\gamma}n_B(\sigma_{3,\gamma\gamma}\omega_{\vb*{q},\gamma}).
\end{align}

The self-energy $\Sigma_{sca}$ is given by
\begin{align}
    \Sigma^{\alpha\beta}_{sca,\vb*{k},i\varepsilon_n} =& -\frac{k_BT}{2N}\sum_{\vb*{q}}\sum_{\gamma,\delta,\delta^\prime}\sum_{m}\tilde{V}_{\vb*{q},\vb*{k}}^{\gamma\alpha\delta}\tilde{V}_{\vb*{k},\vb*{q}}^{\beta\gamma\delta^\prime}\notag\\
    &\times\mathcal{G}_{e,\vb*{q},\gamma}(i\varepsilon_m)\mathcal{G}^{\prime0}_{m,\vb*{q}-\vb*{k},\delta\delta^\prime}(i\varepsilon_m-i\varepsilon_n),
\end{align}
where $\mathcal{G}^{\prime0}_m(i\omega_n)$ is the Fourier transformation of $\mathcal{G}^{\prime0}_m(\tau)$ which is defined as
\begin{align}
    \mathcal{G}^{\prime0}_{m,\vb*{k},\alpha\beta}(\tau) &= -\langle T_\tau \psi_{m,\vb*{k},\alpha}(\tau)\psi_{m,-\vb*{k},\beta}\rangle \notag\\
    &= \sum_{\beta^\prime}\mathcal{G}^0_{m,\vb*{k},\alpha\beta^\prime}(\tau)\sigma_{1,\beta^\prime\beta},
\end{align}
where $\sigma_1$ is $\sigma_x\otimes 1_n$ and we use $\psi_{m,\vb*{k},\alpha}=\psi^\dagger_{m,-\vb*{k},\alpha^\prime}\sigma_{1,\alpha^\prime\alpha}$. Therefore, we obtain
\begin{align}
    \Sigma^{\alpha\beta}_{sca,\vb*{k},i\varepsilon_n} =& -\frac{k_BT}{2N}\sum_{\vb*{q}}\sum_{\gamma,\delta,\delta^\prime}\sum_{m}\tilde{V}_{\vb*{q},\vb*{k}}^{\gamma\alpha\delta}\tilde{V}_{\vb*{k},\vb*{q}}^{\beta\gamma\delta^\prime}\notag\\
     &\times\frac{1}{i\varepsilon_m-\varepsilon_{\vb*{q},\gamma}}\frac{\sigma_{1,\delta\delta^\prime}\sigma_{3,\delta\delta}}{i\varepsilon_m-i\varepsilon_n-\sigma_{3,\delta\delta}\omega_{\vb*{q}-\vb*{k},\delta}}\notag\\
    &=\frac{1}{2N}\sum_{\vb*{q}}\sum_{\gamma,\delta,\delta^\prime}\int_C\frac{dz}{2\pi i}F(z)\tilde{V}_{\vb*{q},\vb*{k}}^{\gamma\alpha\delta}\tilde{V}_{\vb*{k},\vb*{q}}^{\beta\gamma\delta^\prime}
    \notag\\
    &\times\frac{1}{z-\varepsilon_{\vb*{q},\gamma}}\frac{\sigma_{1,\delta\delta^\prime}\sigma_{3,\delta\delta}}{z-i\varepsilon_n-\sigma_{3,\delta\delta}\omega_{\vb*{q}-\vb*{k},\delta}}
\end{align}
where $F(z)=1/(e^{\beta z}+1)$ and $C$ is a contour which encloses the poles of $F(z)$.
By evaluating the integral, we obtain
\begin{align}
    \Sigma^{\alpha\beta}_{sca,\vb*{k},\omega}
    =& \frac{1}{2N}\sum_{\vb*{q}}\sum_{\gamma,\delta,\delta^\prime}\tilde{V}_{\vb*{q},\vb*{k}}^{\gamma\alpha\delta}\tilde{V}_{\vb*{k},\vb*{q}}^{\beta\gamma\delta^\prime}\sigma_{1,\delta\delta^\prime}\sigma_{3,\delta\delta}\notag\\
    &\times\frac{f(\varepsilon_{\vb*{q},\gamma})+n_B(\sigma_{3,\delta\delta}\omega_{\vb*{q}-\vb*{k},\delta})}{\omega-\varepsilon_{\vb*{q},\gamma}+\sigma_{3,\delta\delta}\omega_{\vb*{q}-\vb*{k},\delta}+i\eta}\notag\\
    =&\frac{1}{2N}\sum_{\vb*{q}}\sum_{\gamma,\delta} \notag\\
    &\sigma_{3,\delta\delta}\frac{\tilde{V}_{\vb*{q},\vb*{k}}^{\gamma\alpha\delta}(\tilde{V}_{\vb*{q},\vb*{k}}^{\gamma\beta\delta})^*(f(\varepsilon_{\vb*{q},\gamma})+n_B(\sigma_{3,\delta\delta}\omega_{\vb*{q}-\vb*{k},\delta}))}{\omega-\varepsilon_{\vb*{q},\gamma}+\sigma_{3,\delta\delta}\omega_{\vb*{q}-\vb*{k},\delta}+i\eta}.
\end{align}
Here, we use the relation $\sum_{\delta^\prime}\tilde{V}_{\vb*{k},\vb*{q}}^{\beta\gamma\delta^\prime}\sigma_{1,\delta\delta^\prime}=(\tilde{V}_{\vb*{q},\vb*{k}}^{\gamma\beta\delta})^*$.

The self-energy $\Sigma_{tad}$ is given by
\begin{align}
    \Sigma_{tad,\vb*{k},\omega}^{\alpha\beta}=&\frac{k_BT}{2N}\sum_{\vb*{q}}\sum_{\gamma,\delta,\delta^\prime}\sum_m\tilde{V}_{\vb*{k},\vb*{k}}^{\beta\alpha\delta}\tilde{V}_{\vb*{q},\vb*{q}}^{\gamma\gamma\delta^\prime}\notag\\
    &\times\mathcal{G}_{e,\vb*{q},\gamma}(i\varepsilon_m)\mathcal{G}^{\prime0}_{m,\vb*{0},\delta\delta^\prime}(0)\notag\\
    =&-\frac{1}{2N}\sum_{\vb*{q}}\sum_{\gamma,\delta}\tilde{V}_{\vb*{k},\vb*{k}}^{\beta\alpha\delta}(\tilde{V}_{\vb*{q},\vb*{q}}^{\gamma\gamma\delta})^*\frac{f(\varepsilon_{\vb*{q},\gamma})}{\omega_{\vb*{0},\delta}}.
\end{align}

\section{Self-energy of magnons}\label{appendix:magnon}
\begin{figure}[tb]
    \centering
    \includegraphics[width=\linewidth]{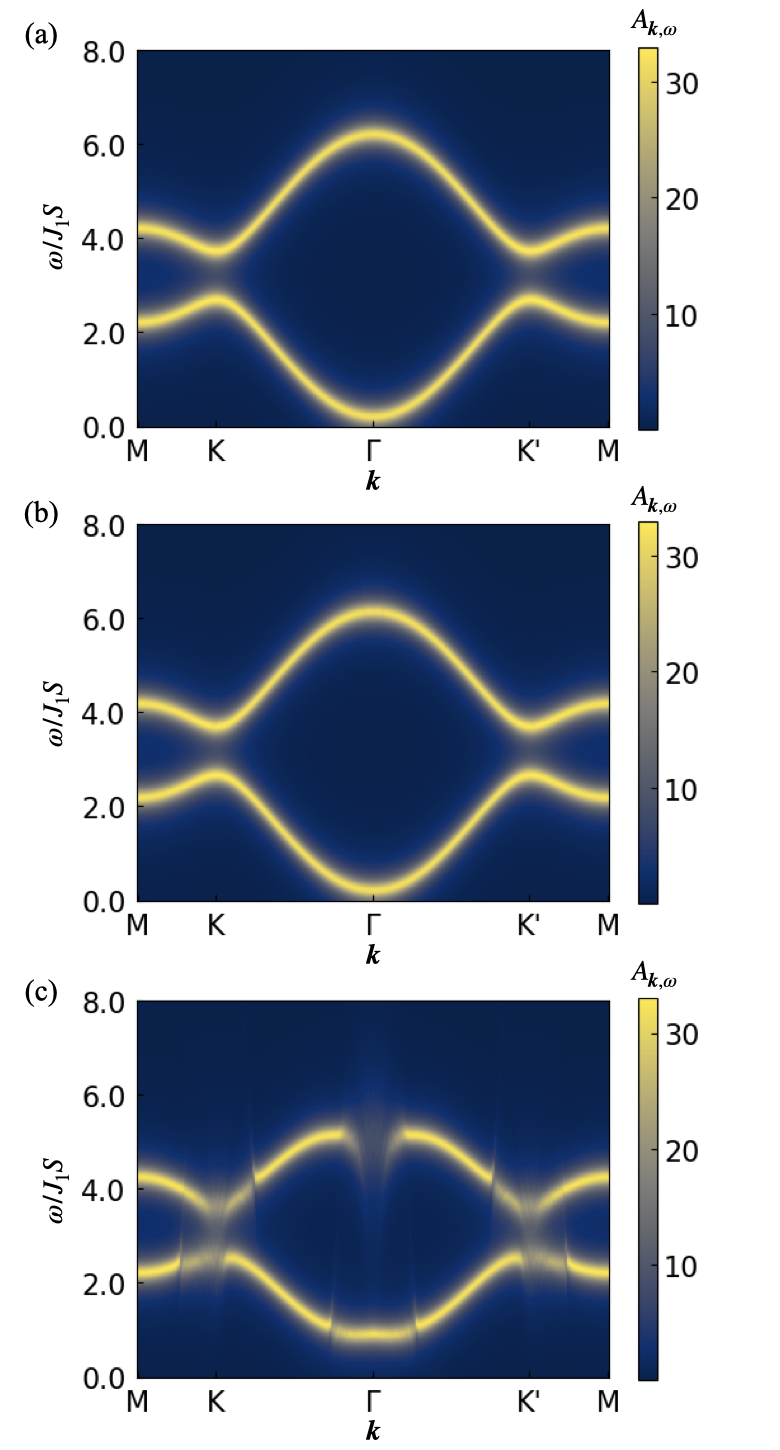}
    \caption{Spectral function of magnons. (a) A spectral function without self-energy along the high-symmetry path. (b) A spectral function with self-energy ($J_{ex}/t_1=0.1$). (c) A spectral function with self-energy ($J_{ex}/t_1=0.5$). We use following parameters: $t_2/t_1=0.0$, $J_1/t_1=0.05$, $J_2/t_1=0.0$, $D/t_1=0.005$, $S=1$, $\Delta_z/t_1=0.005$, $\varepsilon_F/t_1=0.0$ and $\eta/t_1=0.01$} 
    \label{fig:magnon_spectl}
\end{figure}

In the main text, we evaluate the electron self-energy using the unperturbed magnon Green's function. However, the electron-magnon interaction also renormalizes the magnon propagator.
Here, we consider the effect of the electron-magnon interaction for magnons. Similar to the electrons, we evaluate the self-energy and Green's function of magnons. The self-energy from $\mathcal{H}_{\textrm{int}2}$ is given by
\begin{equation}
    \Sigma_{m,cor,\vb*{k},i\omega_n}^{\alpha\beta}=\frac{1}{N}\sum_{\vb*{q}}\sum_{\gamma}\tilde{W}^{\gamma\gamma\alpha\beta}_{\vb*{q},\vb*{q},\vb*{k}}f(\varepsilon_{\vb*{q},\gamma}),
\end{equation}
and self-energy from $\mathcal{H}_{\textrm{int}1}$ is given by
\begin{align}
    \Sigma^{\alpha\beta}_{m,sca,\vb*{k},i\omega_n} =& -\frac{k_BT}{2N}\sum_{\vb*{q}}\sum_{\gamma,\delta}\sum_{m}\tilde{V}_{\vb*{q},\vb*{q}-\vb*{k}}^{\gamma\delta\alpha}(\tilde{V}_{\vb*{q},\vb*{q}-\vb*{k}}^{\gamma\delta\beta})^*\notag\\
    &\times\mathcal{G}_{e,\vb*{q},\gamma}(i\varepsilon_m)\mathcal{G}^{\prime0}_{\vb*{q}-\vb*{k},\delta}(i\varepsilon_m-\sigma_{3,\alpha\alpha}i\omega_n),
\end{align}
and we obtain
\begin{align}
    \Sigma^{\alpha\beta}_{m,sca,\vb*{k},\omega} =& \frac{1}{2N}\sum_{\vb*{q}}\sum_{\gamma,\delta}\sum_{m}\tilde{V}_{\vb*{q},\vb*{q}-\vb*{k}}^{\gamma\delta\alpha}(\tilde{V}_{\vb*{q},\vb*{q}-\vb*{k}}^{\gamma\delta\beta})^*\notag\\
    &\times\frac{f(\varepsilon_{\vb*{q}-\vb*{k},\delta})-f(\varepsilon_{\vb*{q},\gamma})}{\sigma_{3,\alpha\alpha}(\omega+i\delta)-\varepsilon_{\vb*{q},\gamma}+\varepsilon_{\vb*{q}-\vb*{k},\delta}}.
\end{align}
The renormalized Green's function of magnons is defined as
\begin{equation}
    \mathcal{G}_{m,\vb*{k}}(\omega)=(\sigma_{3}(\omega+i\eta)-\omega_{\vb*{k}}-\Sigma_{m,cor,\vb*{k},\omega}-\Sigma_{m,sca,\vb*{k},\omega})^{-1}
\end{equation}
and we evaluate the spectral function as shown in Fig.~\ref{fig:magnon_spectl}. In the absence of the self-energy, the magnon spectrum is given by the unperturbed magnon Green's function as shown in Fig.~\ref{fig:magnon_spectl} (a). When we take into account the self-energy, the magnon spectrum is renormalized as shown in Fig.~\ref{fig:magnon_spectl} (b) and (c). From first-principles, the order of $J_{ex}/t_1$ is $\sim0.1$ ~\cite{Zhang2018StrongHeterostructures,Farooq2019SwitchableHeterostructures,Holmes2020ExchangeHeterostructures}, where the magnon spectrum remains almost unchanged from the non-interacting case as shown in Fig.~\ref{fig:magnon_spectl} (b). When the interaction $J_{ex}$ becomes large, the magnon spectrum is broadened and the energy of the magnon is modulated as shown in Fig.~\ref{fig:magnon_spectl} (c). However, the topological gap around $K$ and $K^\prime$ points is still open and the topological nature of the magnon remains even under the electron-magnon interaction. Thus, the effect of the self-energy of magnons is not essential for the topological nature of the electronic state.

\section{A model of magnetic metal}\label{sec:appendix_model}

In this section, we consider the model of magnetic metal or metal/magnet bilayer systems. We consider a two-dimensional system with a honeycomb lattice structure. The Hamiltonian of the system is given by
\begin{equation}
    \mathcal{H} = \mathcal{H}_{\text{e}} + \mathcal{H}_{\text{S}} + \mathcal{H}_{\text{int}},
\end{equation}
where the electron Hamiltonian $\mathcal{H}_{\text{e}}$ is given by Eq.~\eqref{eq:electron}, the spin Hamiltonian $\mathcal{H}_{\text{S}}$ is given by $\mathcal{H}_{\text{S}_b}$ in Eq.~\eqref{eq:spin_model}, and the interaction Hamiltonian $\mathcal{H}_{\text{int}}$ is given by Eq.~\eqref{eq:interaction}. 
This Hamiltonian is realized in magnetic metals or bilayer systems of nonmagnetic metal and magnetic insulators.

By using the Holstein-Primakoff transformation, we can obtain the magnon Hamiltonian for the bottom layer, which is given by $\mathcal{H}_{\text{m}_b}$ in Eq.~\eqref{eq:magnon}. The interaction between the electron and the magnon is given by
\begin{align}
\mathcal{H}_{\text{int}} = -\sqrt{2S}J_{ex}\sum_{\vb*{k},\vb*{q}}\sum_l(c_{l,\downarrow,\vb*{k}}^\dagger c_{l,\uparrow,\vb*{q}}a_{b,l,\vb*{k}-\vb*{q},}+h.c.) \label{eq:int_appendix}
\end{align}

This model can be decoupled into independent spin-up and spin-down electron systems and the self-energy, spectral function, and effective Hamiltonian of the electron system can be divided into the spin-up and spin-down parts.

\begin{figure}[tbp]
    \centering
    \includegraphics[width=\linewidth]{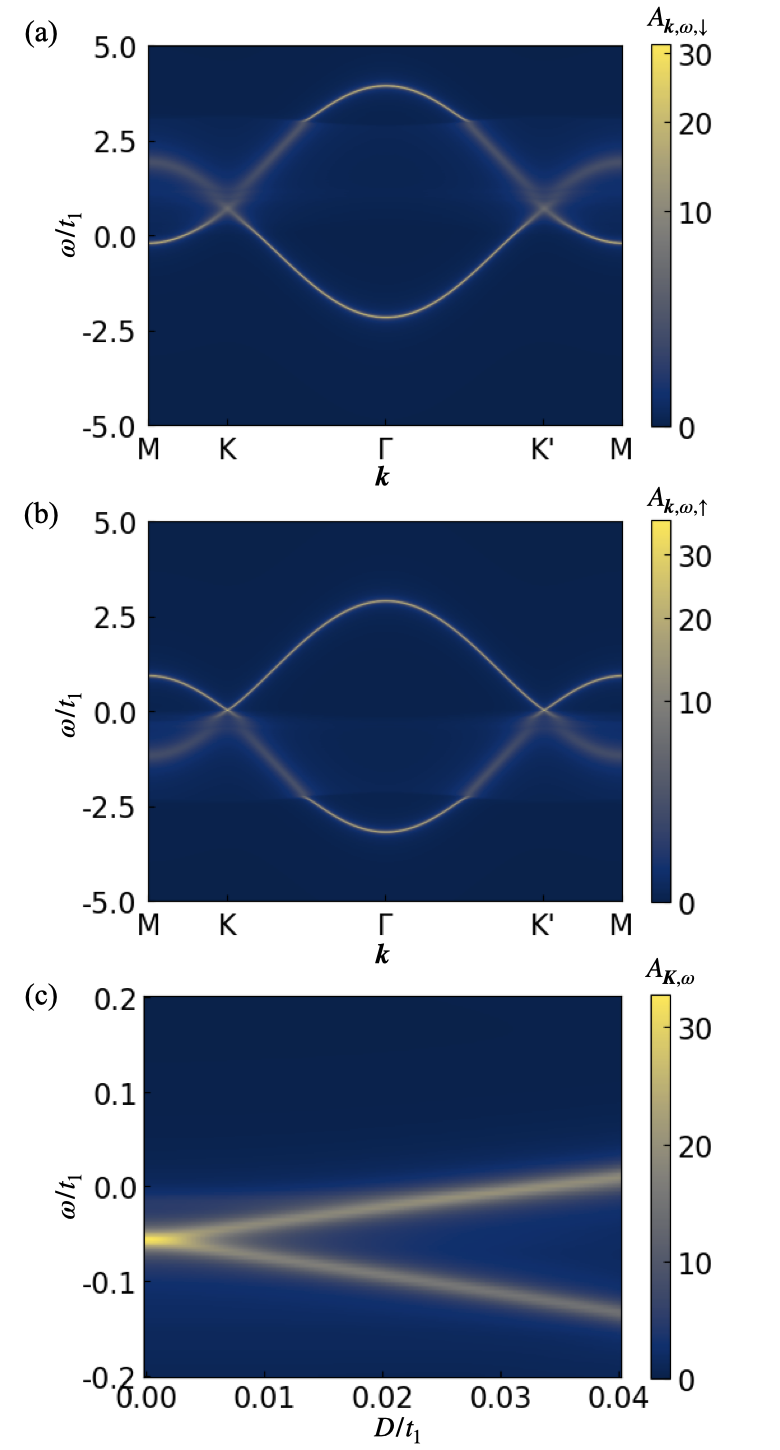}
    \caption{Spectral function of electrons. (a) A spectral function of spin-down electrons along the high-symmetry path($D/t_1=0.0$). (b) A spectral function of spin-up electrons along the high-symmetry path ($D/t_1=0.0$). (c) Spectral function $A_{\vb*{k},\omega}$ at the $K$ point as a function of DM interaction. We use following parameters: $t_2/t_1=0.0$, $J_1/t_1=0.05$, $J_2/t_1=0.01$, $J_{ex}/t_1=0.5$, $S=1$, $\varepsilon_F/t_1=-0.35$, $\Delta_z/t_1=0.005$, $\eta/t_1=0.01$ and $k_BT=0.0$}
    \label{fig:appendix_spectl}
\end{figure}

Figure~\ref{fig:appendix_spectl} shows the result for the spectral function of this model. A finite net magnetization, $\Braket{S^z_i}\neq0$, gives rise to a non-zero exchange field $-J_{ex}\Braket{S^z_i}$. This exchange field splits the spin-up and spin-down bands. Figures ~\ref{fig:appendix_spectl}(a) and (b) show the spectral function of spin-down and spin-up electrons, respectively. Following the discussion in Sec.~\ref{sec:QH}, assuming that the magnon energy is significantly smaller than the electron energy, Eq.~\eqref{eq:self_energy_decoupled} indicates that the imaginary part of the self-energy, $\text{Im}\Sigma(\omega)^{\mu\mu}$, becomes large when there exist electronic states at the energy $\omega$ satisfying $\omega = \varepsilon_{\nu,\vb*{q}}$ ($\nu\neq\mu$) at some momentum $q$. Due to the large imaginary part of the self-energy, lifetime becomes small. Then, the lifetime of the Dirac point of the spin-down band is reduced. However, the Dirac point of the spin-up band around Fermi level has a long lifetime. 
Figure~\ref{fig:appendix_spectl} (c) shows the energy gap at the $K$ point as a function of the DM interaction. In this model, the energy gap at the $K$ point is also induced by the DM interaction.

\begin{figure}[tbp]
    \centering
    \includegraphics[width=\linewidth]{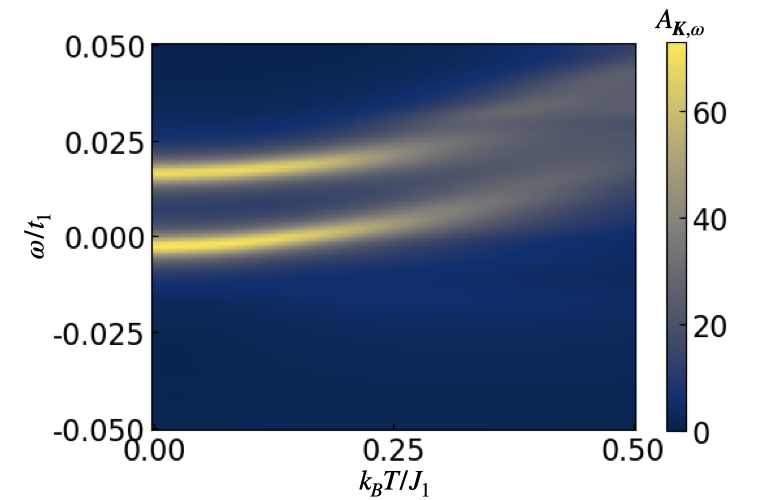}
    \caption{Temperature dependence of spectral function at $K$ point. Two energy bands are up-spin bands and the energy gap corresponds to the topological gap. We use following parameters: $t_2/t_1=0.0$, $J_1/t_1=0.05$, $J_2/t_1=0.01$, $D/t_1=0.035$, $J_{ex}/t_1=0.5$, $S=1$, $\Delta_z/t_1=0.005$, $\varepsilon_F/t_1=-0.35$ and $\eta/t_1=0.003$} 
    \label{appendixfig:T_dep}
\end{figure}

Now, we evaluate the temperature dependence of the topological gap. The self-energy $\Sigma_{\vb*{k},\omega}$ is temperature dependent, and thus the spectral function also depends on the temperature. In particular, the exchange field $-J_{ex}\Braket{S^z_i}$ is modified by the temperature, which causes the spectral function to shift.

\begin{figure}[tbp]
    \centering
    \includegraphics[width=\linewidth]{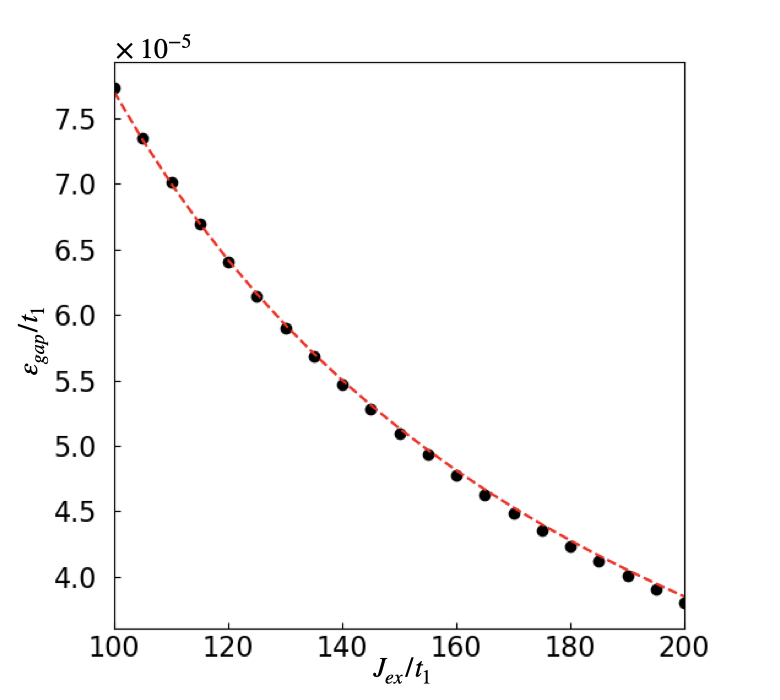}
    \caption{The energy gap at the $K$ point as a function of $J_{ex}/t_1$. Black points are the calculated gap values and the red line is a fitted line. The fitted line is $g(J_{ex}/t_1)=0.0077(J_{ex}/t_1)^{-1}$. 
    We use following parameters: $t_2/t_1=0.0$, $J_1/t_1=0.05$, $J_2/t_1=0.01$, $D/t_1=0.02$, $S=1$, $\Delta_z/t_1=0.005$, $k_BT=0.0$ and $\eta/t_1=0.01$.}
    \label{fig:gap_dep}
\end{figure}

Figure~\ref{appendixfig:T_dep} shows the temperature dependence of the spectral function at the $K$ point around the Fermi level. Two energy bands are the spin-up bands and we focus on the gap between these two bands. At zero temperature, the Fermi level is located inside the topological gap. As the temperature increases, the thermal expectation value of the spin $\Braket{S^z}$ decreases, which in turn reduces the strength of the internal exchange field. Consequently, the spin-up energy band shifts to higher energies and the Fermi level is below the topological gap. This indicates that the energy level of the topological gap can be controlled by the temperature. We assume that the spin system has a ground state spin configuration. Thus, our theory is valid at low temperatures where the magnon picture is applicable.

Finally, we validate the real space picture of the Hall insulator explained as shown in Fig.~\ref{fig:schematic} in Sec.~\ref{sec:intro}.

We now estimate the magnitude of the effective next-nearest-neighbor hopping and the energy scale of the topological gap at zero temperature. 
We assume that $J_{ex}$ is sufficiently larger than the electron hopping $t_1$. In this situation, we consider the effective next-nearest-neighbor hopping for the spin-up electrons. We assume that $\mathcal{H}_{\text{int}1}$ is the perturbation term.
In the ground state, there is no magnon and the electron-magnon interaction creates a magnon with flipping spin of an electron. Then, the virtual intermediate state contains one magnon and a spin-flipped electron. We denote the energy difference between the ground state and the intermediate state as $\Delta E$. The electron and magnon can hop to the next-nearest-neighbor sites by the electron hopping $t_1$ and the magnon hopping $J_1$ and $J_2^\prime e^{i\phi}=J_2+iD$. These hoppings can be described by the tight-binding Hamiltonian of electrons and magnons. Finally, the electron-magnon interaction annihilates the magnon and flips the spin of the electron. This process results in an effective next-nearest-neighbor hopping for the electron. Estimating the order of the effective next-nearest-neighbor hopping $t_\mathrm{eff}$ is given by
\begin{equation}
t_\mathrm{eff}\propto\frac{(J_{ex}\sqrt{S})^2(t_1+J_2^\prime e^{i\phi})(t_1+J)}{(-\Delta E)^3},
\end{equation}
where $J$ is the order of the on-site potential of magnon and $\Delta E$ is the energy difference between the ground state and the intermediate state.
Since we consider the strong-coupling limit ($J_{ex}\gg t_1$), the virtual intermediate state, which contains a spin-flipped electron and one magnon, has an energy of approximately $2J_{ex}S$ above the ground state. Then, the energy difference $\Delta E$ is proportional to $J_{ex}$. Furthermore, we assume that the electron hopping $t_1$ is the dominant term in the intermediate hopping process (i.e., $t_1\gg J,J_2$). Therefore, the order of the imaginary part of the effective next-nearest-neighbor hopping is proportional to $t_1J_2^\prime \sin{\phi}/J_{ex}$.
Since the topological gap is induced by the effective next-nearest-neighbor hopping, the energy scale of the topological gap is also proportional to $t_1J_2^\prime \sin{\phi}/J_{ex}$.
To validate this picture, we calculate the band gap at the $K$ point as a function of $J_{ex}/t_1$ and check whether it is proportional to $(J_{ex}/t_1)^{-1}$. We numerically calculate this dependence using the effective Hamiltonian in Eq.~\eqref{eq:effective_hamiltonian} for the down-spin electrons around the Dirac point as shown in Fig.~\ref{fig:gap_dep}. The results, plotted in Fig.~\ref{fig:gap_dep}, confirm that the gap is indeed proportional to $(J_{ex}/t_1)^{-1}$, validating our real-space picture.

\section{Wilson loop and $Z_2$ index}\label{appendix:Z2}
Here, we show details of evaluation of Wilson loop and the relation between Wilson loop and Wannier charge center.
We consider the Hamiltonian
\begin{equation}
    H=\sum_{a,b}h^{ab}_{\vb*{k}}\ket{a_{\vb*{k}}}\bra{b_{\vb*{k}}},
\end{equation}
where $a$ and $b$ denote the index of the local basis. The Bloch eigenstate of Hamiltonian is given by $\ket{\psi_{n,\vb*{k}}}=\sum_{a}g_{n,a,\vb*{k}}\ket{a_{\vb*{k}}}$ and the periodic part
of the Bloch eigenstate is $\ket{n_{\vb*{k}}}=e^{-i\vb*{k}\cdot r}\ket{\psi_{n,\vb*{k}}}$.
The Berry connection $A(\vb*{k})$ is defined as $A^{nm}_{\vb*{k}}=i\Braket{n_{\vb*{k}}|\partial_{\vb*{k}}|m_{\vb*{k}}}$. We consider the discretized Berry connection of occupied bands along $k_x$ direction and 
\begin{equation}
A^{nm}_{k_{x,j},k_y}=i\frac{\bra{n_{k_{x,j},k_y}}(\ket{m_{k_{x,j}+\delta k,k_y}}-\ket{m_{k_{x,j},k_y}})}{\delta k},
\end{equation}
where $\ket{n}$ and $\ket{m}$ are the periodic part
of the Bloch eigenstate for occupied bands, $k_{x,j}$ is a discretized wave number which satisfies $k_{x,j}=j\delta k$ and $N_x\delta k$ is a reciprocal lattice vector and $N_x$ is the number of unit cells along the $x$ direction in real space.
Therefore,  the Berry connection in discretized wave number can be approximated as
\begin{align}
    e^{-iA_{k_{x,j},k_y}\delta k}&\sim\delta_{n,m}-iA_{k_{x,j},k_y}\delta k\notag\\
    &=\Braket{n_{k_{x,j},k_y}|m_{k_{x,j+1}}} =: F^{n,m}_{j}.
\end{align}
Therefore, Wilson loop for occupied bands is evaluated as
\begin{align}
    W(k_y) &=\prod_{j=0}^{N_x-1}e^{-iA_{k_{x,j},k_y}\delta k}\notag\\
    &\sim\prod_{j=0}^{N_x-1} F_{j,j+1}.
\end{align}
The last equation can be numerically calculated easily. 
The eigenvalue of the $\prod_{j=0}^{N_x-1}F_{j,j+1}$ is given by
\begin{equation}
    \lambda_m^{W}(k_y)=|\lambda_m^W|e^{i\theta_m(k_y)}.
\end{equation}
Now, we show the relation between the Wilson loop and the Wannier center.

The eigenvalues $\lambda_m^W(k_y)$ of $W(k_y)$ are related to the Wannier charge centers $\bar{x}_m(k_y)$ \cite{Yu2011EquivalentConnection}. Specifically, the phases $\theta_m(k_y)$ of the eigenvalues, defined by $\lambda_m^W(k_y) = e^{i\theta_m(k_y)}$, correspond to the $x$-positions of the Wannier charge centers (in units of $2\pi/a$) for the occupied bands within the slice of fixed $k_y$.

\nocite{*}
\bibliography{references}

\end{document}